\documentclass[prd,aps, nofootinbib, preprint, superscriptaddress]{revtex4-1}
\pdfoutput=1
 
\usepackage{amsmath}
\usepackage{amssymb}
\usepackage{epsfig}
\usepackage{graphicx}
\usepackage{subfigure}
\usepackage{slashed}
\usepackage{soul}
\usepackage{url}
\usepackage[normalem]{ulem} 
\usepackage{color}

\newcommand{\of}[1]{\!\left( #1 \right)}
\newcommand{\ofp}[1]{\left( #1 \right)}
\newcommand{\sqof}[1]{\left[ #1 \right]}

\newcommand{\abs}[1]{\left| #1 \right|}

\begin{document}

\title{Cosmological Phase Transitions and their \\ Properties in the NMSSM}

\author{Jonathan Kozaczuk}
\email{jkozaczuk@triumf.ca}\affiliation{TRIUMF, 4004 Wesbrook Mall, Vancouver, BC V6T 2A3, Canada}

\author{Stefano Profumo}
\email{profumo@ucsc.edu}\affiliation{Department of Physics, University of California, 1156 High St., Santa Cruz, CA 95064, USA}\affiliation{Santa Cruz Institute for Particle Physics, Santa Cruz, CA 95064, USA} 

\author{Laurel Stephenson Haskins}
\email{laurelesh@gmail.com}\affiliation{Department of Physics, University of California, 1156 High St., Santa Cruz, CA 95064, USA}\affiliation{Santa Cruz Institute for Particle Physics, Santa Cruz, CA 95064, USA} 

\author{Carroll L. Wainwright}
\email{cwainwri@ucsc.edu}\affiliation{Department of Physics, University of California, 1156 High St., Santa Cruz, CA 95064, USA}\affiliation{Santa Cruz Institute for Particle Physics, Santa Cruz, CA 95064, USA} 

\date{\today}

\begin{abstract}

\noindent We study cosmological phase transitions in the Next-to-Minimal Supersymmetric Standard Model (NMSSM) in light of the Higgs discovery.  We use an effective field theory approach to calculate the finite temperature effective potential, focusing on regions with significant tree-level contributions to the Higgs mass, a viable neutralino dark matter candidate, 1-2 TeV stops, and with the remaining particle spectrum compatible with current LHC searches and results.  The phase transition structure in viable regions of parameter space exhibits a rich phenomenology, potentially giving rise to one- or two-step first-order phase transitions in the singlet and/or $SU(2)$ directions.  We compute several parameters pertaining to the bubble wall profile, including the bubble wall width and $\Delta\beta$ (the variation of the ratio in Higgs vacuum expectation values across the wall).  These quantities can vary significantly across small regions of parameter space and can be promising for successful electroweak baryogenesis.  We estimate the wall velocity microphysically, taking into account the various sources of friction acting on the expanding bubble wall.  Ultra-relativistic solutions to the bubble wall equations of motion typically exist when the electroweak phase transition features substantial supercooling.  For somewhat weaker transitions, the bubble wall instead tends to be sub-luminal and, in fact, likely sub-sonic, suggesting that successful electroweak baryogenesis may indeed occur in regions of the NMSSM compatible with the Higgs discovery.
\noindent \end{abstract}

\maketitle

\section{Introduction}\label{sec:intro}

The origin of the matter-antimatter asymmetry in the Universe remains one of the key open problems at the interface of cosmology and particle physics. While a primordial asymmetry is not observationally ruled out, the relatively small size of the observed asymmetry --- roughly one part in 10 billion --- combined with the success of the inflationary paradigm in diluting away primordial relics, suggest a dynamic origin of the asymmetry. Many such baryogenesis  mechanisms have been proposed, occurring at widely differing points in the history of the Universe, and involving correspondingly different physical scales \cite{baryogenesisreview}.

Among baryogenesis scenarios, those invoking physics at the electroweak phase transition (electroweak baryogenesis, or EWB) have attracted much interest. This is because such models generically require new physics at scales that are within the reaches of current-generation colliders, intensity-frontier precision measurements, and, potentially, planned gravitational wave detectors at the cosmic frontier (for a recent review of electroweak baryogenesis see e.g. \cite{ewbreview}). One framework which, in principal, contains all of the ingredients  necessary for successful electroweak baryogenesis  is the minimal supersymmetric extension of the Standard Model, or MSSM \cite{mssmewb}.

The discovery of a Higgs particle at the Large Hadron Collider (LHC), and the lack of any signals for supersymmetric particles, have started to put significant pressure on the corner of MSSM parameter space compatible with EWB \cite{Carena:2012np}. In particular, it was shown in a recent analysis \cite{Carena:2012np} that requiring both a strongly first-order phase transition and a Higgs mass around 125 GeV forces one of the two stop squark masses to be in the $10^6$ TeV range, with the other one around 120 GeV.  This in turn requires a light neutralino, with a mass lower than 60 GeV, contributing to the invisible decay width of the Higgs and thereby reducing the enhancement in gluon-gluon  fusion Higgs production from the light stop \cite{Menon:2009mz}. Additionally, searches for stop squarks at the LHC have all but ruled out the light stop scenario in the MSSM directly \cite{Krizka:2012ah, Delgado:2012eu}, albeit with some possible loopholes.  Apart from the requirement of a strongly first order phase transition, the CP-violating sources in the MSSM are also highly constrained by current limits from electric dipole moment experiments \cite{Cirigliano:2009yd, Kozaczuk:2012xv, Kozaczuk:2012vx}. In light of these recent developments, MSSM EWB appears quite unnatural at best, if not plainly contrived.  

An alternative supersymmetric framework for electroweak baryogenesis is one in which the MSSM Higgs sector (consisting of two Higgs doublets) is enlarged to include an additional singlet superfield --- a setup known as the next-to-minimal supersymmetric extension to the Standard Model, or NMSSM. It has long been appreciated \cite{Pietroni:1992in,  Davies:1996qn} that tree-level cubic terms in the Higgs potential might provide a strongly first-order electroweak phase transition, relaxing the model-building challenges one faces in the MSSM.  Nevertheless, relatively little is known about the details of how the phase transition proceeds in this case.  Many previous studies of the NMSSM and related non-minimal SUSY models have exclusively considered one-step electroweak phase transitions \cite{Pietroni:1992in, Davies:1996qn, Huber:2000mg, oai:arXiv.org:hep-ph/0404184, Ham:2004nv, Huber:2006ma, Huber:2006wf, Ham:2007wu, LightDM_EWPT, Kozaczuk:2013fga,  Balazs:2013cia}, while others have considered the possibility of a more complicated pattern of symmetry breaking \cite{Funakubo:2005pu}, albeit prior to the Higgs discovery and with several assumptions about the spectrum that are no longer justified.  Also lacking are detailed predictions for the quantities pertaining to the bubble wall profile and which enter the calculation of the baryon asymmetry in the context of EWB.  To the best of our knowledge, amongst the previously mentioned studies only Ref.~\cite{Huber:2000mg} attempts this task. However, they considered a slightly different scenario, and, more critically, did not have knowledge of the Standard Model Higgs discovery.  A major aim of our study is to bridge this gap.

The baryon asymmetry produced during electroweak baryogenesis often depends crucially on the $SU(2)$ and singlet wall thickness between the broken and unbroken electroweak phase ($L_w$ and $L_s$), the relative variation of the neutral $SU(2)$ Higgs fields across the wall (a quantity dubbed $\Delta\beta$), and the velocity of the wall relative to the plasma ($v_w$). Although there are some notable exceptions \cite{Cline:2000kb, Konstandin:2005cd}, in many cases the resulting baryon asymmetry is directly proportional to $\Delta\beta$ and inversely proportional to $L_w$ \cite{Lee:2004we, Kozaczuk:2011vr}.  The dependence on the wall velocity is more complicated \cite{Lee:2004we, Kozaczuk:2011vr}: if $v_w$ is too small, the baryon asymmetry will be diluted by interactions with the plasma in front of the bubble wall, while for quickly moving walls, the $SU(2)$ sphalerons do not have sufficient time to act on the diffusing chiral current, leading again to a suppression of the asymmetry.   In the literature, typical ranges for $L_w$ and $\Delta \beta$ have been given for the case of the MSSM \cite{Moreno:1998bq} and the general (non-scale-invariant) NMSSM \cite{Huber:2000mg}, but no detailed studies exist taking into account the crucial requirement of a 125 GeV SM-like Higgs.  Little has been discussed in the literature about what drives these quantities and how sensitive their values are to the underlying input parameters of the theory.  On the other hand, the wall velocity and the related issue of runaway bubble walls (i.e. bubbles which accelerate indefinitely) have not yet been rigorously addressed in the NMSSM.

The present study also addresses a few technical aspects of the calculation of the electroweak phase transition that had previously been glossed over in the NMSSM. Most importantly, to deal with potentially large logarithms arising in the one-loop effective potential, we integrate out both stop squarks (which we assume are heavy) and work in an effective field theory with two Higgs doublets, a singlet, neutralinos/charginos, and the remaining SM spectrum.  We have implemented this setup in the widely--used {\tt CosmoTransitions} software package \cite{Wainwright:2011kj} and detail our set-up for use in future studies.

Our approach will be to focus on a well-defined corner of the NMSSM parameter space which is both phenomenologically viable and demonstrative of a broad variety of outcomes for the patterns of electroweak symmetry breaking. By ``viable'' we mean that every one of the points we consider features a Higgs sector entirely compatible with results from the LHC, a sparticle spectrum compatible with LHC searches, a lightest neutralino with a thermal relic abundance matching the observed dark matter density, and with direct and indirect detection rates in accordance with current limits.

Focusing on a specific region of the NMSSM parameter space, we produced a set of points exhibiting several different patterns of electroweak symmetry breaking. For each point, we calculated the various phase transition properties, including the relevant friction coefficients needed for a detailed estimate of the bubble wall velocity, as well as the predicted spectrum of gravitational waves from bubble collisions in the early universe.   One of our key findings is that even this small region of parameter space can exhibit a broad variety of outcomes for the quantities relevant for EWB. Parameters pertaining to the bubble wall profile and expansion, which are critical inputs for computing the baryon asymmetry in EWB, can vary by up to an order of magnitude, even across the small slice of parameter space near our benchmarks, and take on favorable values for successful EWB. For example, the $SU(2)$ wall width varies in the broad range $1\lesssim L_wT\lesssim 20$; the parameter $\Delta\beta$ also spans a large range of values, $0.01\lesssim\Delta\beta\lesssim0.5$; the bubble wall velocity is found to be roughly $\sim \mathcal{O}(0.01-0.1)$ for sub-sonic bubbles.  We do not find any detectable level of gravitational wave emission for any of the points considered.

The rest of this manuscript is organized as follows. Sec.~\ref{sec:NMSSM} introduces the model under consideration (the scale-invariant NMSSM), the associated effective potential, and our computational strategy.  In particular, we explain how we deal with large logarithms, and detail how we implement the effective potential in \texttt{CosmoTransitions}. Sec.~\ref{sec:params} presents the parameter space we consider, focusing on the phenomenology of the relevant particle sectors (neutralinos/charginos, sfermions, Higgses). Subsequently, Sec.~\ref{sec:EWPT} describes how we study the electroweak phase transition and its associated properties, and presents our key findings. The computation of the wall velocity is then given in Sec.~\ref{sec:vw}, including details about the setup used to solve the wall's equation of motion, the relevant microphysical sources of friction, and their computation. Our summary and conclusions are given in Sec.~\ref{sec:conc}, just before two appendices: one (App.~\ref{ap:RGEs}) lists the relevant renormalization group equations for the parameters entering the effective potential, and the other (App.~\ref{ap:rates}) provides the matrix elements used to estimate the various friction coefficients governing the expansion of the bubble wall in the plasma.

\section{The Scale-Invariant NMSSM and Effective Field Theory}\label{sec:NMSSM}

\subsection{Overview of  the Strategy}

Before delving into the details of our analysis, we outline the key steps comprising our strategy for studying phase transitions in the NMSSM:
\begin{enumerate}
\item Begin by choosing NMSSM parameters at $M_{\widetilde{t}}$, the scale set by the stop squark masses.  These can be fed into a spectrum calculator (we use \texttt{NMSSMTools} \cite{NMSSMTools}, as discussed further below) for a precise determination of the spectrum to check against phenomenological constraints.
\item Integrate out the stop squarks from the spectrum, since they are assumed to be heavy.  The resulting theory contains two complex Higgs doublets, Higgsinos, winos, a complex singlet, singlino, and a bino, along with the rest of the Standard Model spectrum.  This results in a (non-supersymmetric) two Higgs doublet plus singlet (2HD+S) potential for the scalar fields.
\item Match the NMSSM parameters at $M_{\widetilde{t}}$ onto the parameters of the 2HD+S potential, including the relevant threshold corrections.
\item Solve the renormalization group equations (RGEs) for the various quartic and dimensionful parameters in the potential to evolve them from $M_{\widetilde{t}}$ to $m_t$.  We will do so approximately using a fixed order approximation, but this can also be done exactly.
\item Include the 1-loop zero temperature contributions to the potential.  Impose counterterms to minimize the dependence of the effective potential on the renormalization scale $\Lambda$.
\item Add the finite temperature contribution to the potential.  The full 1-loop effective potential can then be fed into \texttt{CosmoTransitions} \cite{Wainwright:2011kj} to compute the phase transition properties, as discussed in more detail in Sec.~\ref{sec:PT}.
\end{enumerate}

Readers not concerned with the details of the above procedure should proceed directly to our discussion of the particle spectra in Sec.~\ref{sec:params}.  The remainder of this Section is devoted to elucidating the aforementioned steps of our analysis.
  
\subsection{The Model}
Our starting point is the so-called scale-invariant NMSSM.  In this incarnation, a $\mathbb{Z}_3$ symmetry forbids dimensionful parameters in the superpotential $W$, which is here given by
\begin{equation} \label{eq:W}
W=W_{\rm MSSM}|_{\mu=0}+\lambda \widehat{S}\widehat{H}_u\cdot \widehat{H}_d + \frac{\kappa}{3}\widehat{S}^3.
\end{equation}
The hatted quantities above represent the chiral $SU(2)$ superfields $\widehat{H}_u=\left(\widehat{H}_u^+,\widehat{H}_u^0\right)$,  $\widehat{H}_d=\left(\widehat{H}_d^0,\widehat{H}_d^-\right)$, while $\widehat{S}$ is a gauge singlet chiral superfield. The dot represents the usual anti-symmetric $SU(2)$ product.  Supersymmetry is softly broken, with
\begin{equation}
\Delta V_{soft}=m^2_{H_u} \left|H_u\right|^2+m_{H_d}^2\left|H_d\right|^2+m_{S}^2\left|S\right|^2+\lambda A_{\lambda}H_u\cdot H_d S + \frac{1}{3}\kappa A_{\kappa}S^3.
\end{equation}  The tree-level potential is then given by
\begin{equation}
\begin{aligned}
V=&\frac{g^2}{4}\left(\left|H_u^0\right|^2+\left|H_u^+\right|^2-\left|H_d^0\right|^2-\left| H_d^-\right|^2\right)^2+\frac{g_2^2}{2}\left|H_u^+H_d^{0*}+H_u^0H_d^{-*}\right|^2 +\Delta V_{\rm soft} +\sum_i \left|F_i\right|^2,\\
\end{aligned}
\end{equation}
where $g^2\equiv (g_1^2+g_2^2)/2$, $g_1$ and $g_2$ denote the $U(1)$ and $SU(2)$ gauge couplings, respectively, and the sum over the $F$--terms is over $H_{u,d}^0$, $S$, with $F\equiv\partial W/\partial \phi_i$.  For details regarding the rest of the spectrum, we refer the Reader to Refs.~\cite{Kozaczuk:2013fga, NMSSMTools}, whose notation and conventions we follow here unless otherwise stated (see also Ref.~\cite{Ellwanger:2009dp} for a more detailed review).

Note that, because the superpotential respects a discrete $\mathbb{Z}_3$ symmetry, this model will lead to dangerous domain wall formation in the early universe.  This has been long appreciated and can be avoided e.g. by allowing for non-renormalizable operators which break the symmetry explicitly at some high scale. These operators will have no discernable effect on the electroweak physics we are interested in.  However, in this case, one must still ensure that the new operators do not induce unacceptably large tadpole terms for the singlet.  This can be accomplished by imposing additional discrete symmetries to forbid the dangerous operators at tree-level.  For further discussion we refer the Reader to Ref.~\cite{Ellwanger:2009dp} and the references therein.

\subsection{NMSSM Effective Potential} \label{sec:NMSSM_Veff}
Anticipating electroweak symmetry breaking, we write the Higgs and singlet fields as
\begin{align}
H_u = \frac{1}{\sqrt{2}} \begin{pmatrix} 0\\h_u\end{pmatrix};\;
H_d = \frac{1}{\sqrt{2}} \begin{pmatrix}  h_d\\0\end{pmatrix};\;
S = \frac{1}{\sqrt{2}} s.
\end{align}
In terms of these fields, the portion of the tree-level potential relevant for the electroweak phase transition becomes
\begin{multline} \label{eq:NMSSMTreelevel}
V_0(h_u, h_d, s) =
\frac{1}{32}(g_1^2+g_2^2) \left( h_u^2 - h_d^2\right)^2 + \frac{1}{4} \kappa^2 s^{4} - \frac{1}{2} \lambda \kappa s^2 h_u h_d + \frac{1}{4} \lambda^2 \left(h_d^2 h_u^2 + s^2 \left(h_d^2 + h_u^2\right)\right) \\
 + \frac{\sqrt{2}}{6}  \kappa A_{\kappa}  s^3 - \frac{\sqrt{2}}{2} \lambda A_{\lambda}   s h_u h_d 
+ \frac{1}{2} m_d^2 h_d^2  + \frac{1}{2} m_u^2 h_u^2 + \frac{1}{2} m_s^2 s^2.
\end{multline}
The vacuum expectation values (VEVs) of the scalar fields are assumed to be real at all temperatures. We ensure that the potential is meta-stable in the charged and imaginary directions (that is, its Hessian matrix has only positive eigenvalues), but we do not check for the existence of or tunneling to charged vacua.  After electroweak symmetry breaking, and at zero temperature in the physical vacuum, we define $\langle h_u\rangle\equiv v_u$, $\langle h_d\rangle \equiv v_d$, and $\langle s\rangle \equiv v_s$ as usual.  Note that the non-zero vacuum expectation value of $s$ generates an effective $\mu$ term in the superpotential Eq.~\ref{eq:W}, $\mu \equiv \lambda v_s/\sqrt{2}$.

The zero temperature potential in Eq.~\ref{eq:NMSSMTreelevel} receives quantum corrections from all fields which couple to $h_{u,d}$ and $s$. These corrections are given by the well-known Coleman-Weinberg expression
\begin{align} \label{eq:1loop}
V_1(T\!=\!0) = \sum_i \frac{\pm n_i}{64\pi^2} m_i^4 \left[\log\left(\frac{m_i^2}{\Lambda^2}\right) -c_i\right].
\end{align}
Here $m_i^2$ are the field-dependent mass-squared values, $n_i$ are their associated number of degrees of freedom, and $\Lambda$ is the renormalization scale.  The constants $c_i$ depend on the renormalization scheme.  We choose to work in the $\overline{MS}$ prescription, whereby $c=\tfrac{1}{2}$ for the transverse polarizations of gauge bosons, while $c=\tfrac{3}{2}$ for their longitudinal polarizations and for all other particles. The plus and minus signs in Eq.~\ref{eq:1loop} are for bosons and fermions, respectively. The sum over the relevant particles $i$ includes all Standard Model particles (we ignore fermions lighter than the bottom quark since their Yukawa couplings are small), the physical Higgs and other scalar particles, their associated Goldstone bosons, the neutralinos and the charginos. We work in Landau gauge so that the ghost bosons decouple and need not be included in the spectrum. 

Note that the one-loop potential contains explicit gauge-dependence \cite{Patel:2011th, Wainwright:2011qy, Wainwright:2012zn, Garny:2012cg}, which cancels against the implicit gauge-dependence of the VEVs at every order in $\hbar$. As is common practice, we do not consider the effects of the implicit gauge-dependence, and so our results will contain gauge artifacts.  However, our primary purpose in examining the effective potential is to estimate whether or not a first-order phase transition is possible and to infer its general properties in comparison with e.g. the MSSM, for which purpose a calculation with gauge-dependence is acceptable.   Additionally, since the transitions are driven primarily by the singlet contributions to the potential, Landau gauge would appear to be a reasonable choice, since the gauge-dependent contribution to the finite temperature effective potential in this gauge is small.  Defining a gauge-independent version of the one-loop effective potential including the full bosonic thermal contribution is an open problem and is beyond the scope of the present study.  

We will be primarily interested in regions of the NMSSM with moderately heavy ($\sim 1-2$ TeV) stops.  The stops enter the 1-loop zero-temperature effective potential through Eq.~\ref{eq:1loop}.   The top Yukawa coupling  controls both the coupling of the top quarks and stop squarks to the Higgs fields, and since there is a large hierarchy in scale between the tops and stops, large logarithms will inevitably arise from a naive application of Eq.~\ref{eq:1loop}.  This situation is quite unacceptable for calculating the phase transition properties using the one-loop potential, since the large differences between tree-level and one-loop expressions signify a substantial dependence on the renormalization scale $\Lambda$ and the need to go to higher orders in the loop expansion.  To avoid this issue, we employ an effective field theory approach, which we describe below.  It involves integrating out the stops and working in the resulting model with two Higgs doublets, a singlet, neutralinos/charginos, and the rest of the SM spectrum with a renormalization scale near the electroweak scale.

\subsection{Dealing with Large Logarithms}

In keeping with the usual convention, we will use as inputs the NMSSM parameters defined at the stop mass scale, $M_{\widetilde{t}}$.  To avoid the large logarithms arising from the stops in Eq.~\ref{eq:1loop}, we will integrate the stops out at $M_{\widetilde{t}}$ (see e.g. Refs.~\cite{  Haber:1993an, Carena:1995wu, Carena:2000yi, Carena:2008rt, Draper:2013oza}). Below this scale, the tree-level potential can be described by the following general two Higgs doublet + singlet (2HD+S) potential \cite{  Elliott:1993ex, Elliott:1993uc, Elliott:1993bs}
\begin{equation}\label{eq:2HDS_potential}
\begin{aligned}
V_0 =& \frac{1}{2}\lambda_1 \left| H_d\right|^4 + \frac{1}{2}\lambda_2 \left|H_u\right|^4+\left(\lambda_3+\lambda_4\right) \left|H_d\right|^2\left|H_u\right|^2 -\lambda_4 \left|H_u^{\dagger} H_d\right|^2 +\lambda_5 \left|S\right|^2\left|H_d\right|^2 \\
&+\lambda_6 \left|S\right|^2\left|H_u\right|^2 +\lambda_7\left(S^{*2}H_d\cdot H_u + h.c.\right) + \lambda_8 \left|S\right|^4 + m_1^2\left|H_d\right|^2+m_2^2\left|H_u\right|^2+m_3^2 \left|S\right|^2\\
&-m_4 \left(H_d\cdot H_u S+h.c.\right)-\frac{1}{3} m_5 \left(S^3 +h.c.\right).
\end{aligned}
\end{equation}
Comparing the above potential with Eq.~\ref{eq:NMSSMTreelevel}, at the scale $\Lambda = M_{\widetilde{t}}$ one can define
\begin{equation}\label{eq:2HDS_params}
\begin{aligned}
\lambda_1^0 = \lambda_2^0 = \frac{1}{4}(g_1^2+g_2^2), \hspace{1cm} \lambda_3^0 = &\frac{1}{4}(g_2^2-g_1^2), \hspace{1cm}  \lambda_4^0 = \lambda^2-\frac{1}{2} g_2^2 \\
\lambda_5^0=\lambda_6^0=\lambda^2, \hspace{1cm} \lambda_7^0 &= -\lambda \kappa, \hspace{1cm} \lambda_8^0 = \kappa^2.
\end{aligned}
\end{equation}
Similarly, for the mass terms, we have
\begin{equation}
m_4^0=\lambda A_\lambda \hspace{1cm} m_5^0= -\kappa A_{\kappa}.
\end{equation}

Integrating out the stops results in threshold corrections to the parameters in Eq.~\ref{eq:2HDS_params}.  Keeping only renormalizable terms in the potential, the only relevant threshold correction at $M_{\widetilde{t}}$ is to the $h_u$ quartic,
\begin{equation}
\Delta \lambda_2 = \frac{3 y_t^4 A_t^2}{8\pi^2 M_{\widetilde{t}}^2 } \left(1 - \frac{A_t^2 }{12 M_{\widetilde{t}}^2} \right).
\end{equation}
Then, at the scale $M_{\widetilde{t}}$, the parameters in Eq.~\ref{eq:2HDS_potential} are given by 
\begin{equation}
\begin{aligned}
\lambda_i&=\lambda_i^0 +\Delta \lambda_i\\
m_i&=m_i^0
\end{aligned}
\end{equation}
and where $\Delta\lambda_i=0$ for $i\neq 2$.

Ultimately, in analyzing the phase transition structure of the theory using the effective potential, we would like to work at a renormalization scale $\Lambda \sim m_{t}$ to reduce the logarithmic contribution from the top quark in the physical minimum.  To do so requires the various parameters at the scale $\Lambda$, which can be obtained by solving the relevant renormalization group equations.  The most important contributions to the RGEs are those from the top quarks, gauge bosons, Higgs and singlet bosons, Higgsinos, and singlinos.  We do not include the gaugino contributions, since in the benchmark points we study the wino is always rather heavy and the bino does not significantly affect the running.  We list the relevant RGEs, along with more details about which contributions we consider, in Appendix~\ref{ap:RGEs}.

A complete resummation of the large logarithms requires solving the full set RGEs.  However, for our purposes it is sufficient to consider the lowest order solutions, given by
\begin{equation}\label{eq:fixed_order}
\begin{aligned}
&\lambda_i(m_t) \simeq \lambda_i(M_{\widetilde{t}})-\beta_{\lambda_i} \log\frac{M_{\widetilde{t}}^2}{m_t^2}\\
&m_i(m_t)\simeq m_i(M_{\widetilde{t}})-\beta_{m_i} \log\frac{M_{\widetilde{t}}^2}{m_t^2}.
\end{aligned}
\end{equation}  
The above approximation corresponds to keeping only the first term in the loop expansion.  With the parameters defined at $m_t$, one can take derivatives of the effective potential of Eq.~\ref{eq:2HDS_potential} and obtain the mass matrices for the various Higgs bosons.  We will denote the lightest CP-even (odd) Higgs mass eigenstate as $h_s$ ($a_s$), since it is very singlet-like for all of our benchmarks.  The second-lightest CP-even Higgs mass eigenstate is denoted by $h$ and will be Standard Model-like for all points considered. We use $v=246$ GeV, as well as the values of $\tan \beta$ and $v_s$ (obtained from $\mu$ and $\lambda$) as input parameters at $\Lambda=m_t$ instead of $m_{1,2,3}$, solving for the latter by minimizing the tree-level potential.  We find that the above procedure yields good agreement with more complete calculations, the masses and mixing matrix entries for the scalars and pseudoscalars falling within a few percent of those calculated using \texttt{NMSSMTools} \cite{NMSSMTools}.  For our purposes, this is sufficient and it encapsulates the sizable corrections from the stop sector to the SM-like Higgs mass, as well as the other dominant 1-loop contributions.

The one-loop effective potential given by Eq.~\ref{eq:1loop} has non-zero derivatives at the tree-level minimum, so the one-loop and tree-level minima do not coincide.  We follow the strategy of Ref.~\cite{Cline:2011mm} and employ counterterms to cancel the one-loop effects, ensuring that the electroweak minimum of the one-loop plus counterterm-corrected effective potential is the same as that given by the parameters $\beta$, $\mu$, and $\lambda$. The counterterms can be written as
\begin{equation}
V_{ct} = \delta m_1^2 |H_d|^2 + \delta m_2^2 |H_u|^2 + \delta m_3^2 |S|^2.
\end{equation}
In principle, one could also impose counterterms to ensure that the masses do not change between the tree-level and one-loop potentials.  However, we find that the masses at one-loop do not differ drastically from their tree-level counterparts, and since it is the tree-level masses that enter the one-loop finite temperature contribution to the effective potential, we do not include these terms.

Finally, using the improved effective potential, we can compute the various masses for the particles and use them to calculate the 1-loop zero-temperature effective potential via Eq.~\ref{eq:1loop}. The full one-loop contribution at finite-temperature is given by
\begin{align}\label{eq:finiteT}
V_1(T\!>\!0) = V_1(T\!=\!0) + \frac{T^2}{2\pi^2} \sum_i n_i J_\pm \left(\frac{m_i^2}{T^2}\right), 
\end{align}
where
\begin{align}
J_\pm(x^2) \equiv \pm \int_0^\infty dy \; y^2 \log\left(1 \mp e^{-\sqrt{y^2+x^2}}\right)
\end{align}
and again the upper (lower) signs correspond to bosons (fermions). The sum includes all of the same particles included in Eq.~\ref{eq:1loop}.  At high temperature, the validity of the perturbative expansion of the effective potential breaks down. Quadratically divergent contributions from non-zero Matsubara modes must be re-summed through inclusion of thermal masses in the one-loop propagators \cite{gross1981,parwani1992}. This amounts to adding thermal masses to the longitudinal gauge boson degrees of freedom and to all of the scalars.  That is, the bosonic mass matrices $M_{ij}^2$, from which the individual eigenvalues $m_i^2$ are calculated, receive extra thermal mass contributions $M_{ij}^2 \rightarrow M_{ij}^2 + \Pi_{ij} T^2$. The thermally corrected eigenvalues are then re-input into Eq.~\ref{eq:finiteT}, yielding the re-summed finite-temperature effective potential.

The full one-loop effective potential at finite temperature is then given by 
\begin{align}
V(h_u, h_d, s, T) = V_0(h_u, h_d, s) + V_1(T\!=\!0) + \frac{T^2}{2\pi^2} \sum_i n_i J_\pm \left(\frac{m_i^2}{T^2}\right)+V_{ct}
\end{align}
where the masses $m_i^2$ are field-dependent (calculated from the 2HD+S potential with counterterms) and include thermal mass corrections.  This potential can then be used to determine the phase structure of phenomenologically viable NMSSM parameter space points.

We emphasize that the application of this approach to the NMSSM is a novel feature of our analysis and allows for a correct treatment of the stop sector in studying phase transitions close to the electroweak scale.  For a similar strategy in the MSSM at two loops, see Ref.~\cite{Carena:2008rt}.

With our strategy laid out, we now turn to the parameter space we wish to explore.

\section{The Parameter Space}\label{sec:params}

Even in its $\mathbb{Z}_3$ symmetric incarnation, the NMSSM parameter space is very large.  Instead of performing numerical scans over all of the parameter space, our approach is to study the phase transition properties in a particular region of the NMSSM motivated by Higgs physics, SUSY searches, dark matter constraints, and naturalness arguments.  As we will see, even in the small parametric window we consider, the phase structure exhibits a rich phenomenology. 

Most recent studies \cite{LightDM_EWPT, Balazs:2013cia} have focused on regions with very light Higgs and neutralino states.  In light of the Higgs discovery and the non-observation of non-SM particles at the LHC, we will instead consider parameter space which accommodates:
\begin{itemize}
\item A significant tree-level contribution to the SM-like Higgs mass.  This corresponds to regions with sizable $\lambda$ and low $\tan\beta$.  Note that large values of $\lambda$ are not necessarily required for a strongly first order EWPT \cite{LightDM_EWPT}, but they do reduce the fine-tuning required to obtain the correct Higgs mass \cite{Natural_1, Natural_2, Natural_3}.
\item Very little mixing between the singlet-like and SM-like Higgses, the latter in good agreement with LHC observations.
\item Moderately heavy stops and other sfermions, in the 1--2 TeV range.  This is motivated by naturalness arguments \cite{Natural_1, Natural_2, Natural_3} and the non-observation of superpartners at the LHC.
\item A viable neutralino dark matter candidate which saturates the observed relic abundance and is compatible with direct- and indirect-detection experiments.
\item Higgs and chargino/neutralino spectra compatible with current LHC limits.
\end{itemize} 
Additionally, we will focus on regions with relatively small $\kappa \in [0.1,0.15]$.  This is appealing from the standpoint of electroweak baryogenesis, since $\kappa$ governs the quartic couplings involving the singlet, and a smaller quartic coupling tends to strengthen the phase transition along the corresponding field direction (this behavior is familiar from the SM case \cite{QuirosReview}).  Taken together with our other parametric choices, this results in the singlet-like CP-even Higgs state being lighter than the SM-like Higgs for the points we consider.  We will comment further on this feature below.

Our study will be centered around several representative benchmark points in line with the considerations outlined above.  The details of the benchmarks and their associated phenomenology are given in Table~\ref{tab:bm} below.  These points are in good agreement with all relevant experimental observations, involve relatively low fine-tuning, and will be shown to exhibit dramatically different symmetry breaking patterns in the early universe.  Note that the spectra in Table~\ref{tab:bm} are quite insensitive to changes in $A_{\kappa}$.  To a good approximation, varying $A_{\kappa}$ amounts simply to varying the singlet-like Higgs masses.  This can be seen by comparing the spectra of BM 1 and 2, or BM 3 and 4.  Both pairs feature an identical choice for the remaining parameters, with the individual points differing only in their values for $A_{\kappa}$.  To illustrate the variation of the phase transition properties around the benchmark points, we will vary $A_{\kappa}$ (and hence $m_{h_s}$) around the values listed in Table~\ref{tab:bm}.  For this purpose we define three sets of points, Sets I, II, III, corresponding to the parameter values (except $A_{\kappa}$) for BM 1/2, BM 3/4, and BM 5.  We will scan over $A_{\kappa}$ for these three sets in Sec.~\ref{sec:EWPT}.
 
  \begin{table}[!tc]
\centering

 \begin{tabular}{c|   c  c  c  c  c }
&\multicolumn{2}{c}{Set I}&\multicolumn{2}{c}{Set II}&\multicolumn{1}{c}{Set III}\\
&BM 1  &BM 2 & BM 3 & BM 4 & BM 5  \\
\hline
$\lambda$ &0.63 & 0.63 &0.6&  0.6 & 0.61\\
$\kappa$ & 0.12 &0.12 &0.13  & 0.13 &  0.12  \\
$A_{\lambda}$ [GeV] &  335 & 335 & 350 & 350  & 360 \\
$A_{\kappa}$ [GeV] & -90 & -129 & -56 & -79 & -154 \\
$\tan \beta$ &1.5 & 1.5 & 1.7 & 1.7 & 1.6 \\
$\mu$ [GeV]  &180 & 180 & 180 & 180 & 190 \\
$M_1$ [GeV] &-100.0 & -100.0 & -103.5 & -103.5 & -102.0 \\
$M_{\widetilde{Q}_3}$ = $M_{\widetilde{U}_3}$ [TeV] &  1.0 & 1.0 & 1.5 & 1.5 &1.2  \\
$A_t$ [GeV] & 400 & 400 & 1500 & 1500 & 1200  \\
\hline
\hline
$m_{h}$ [GeV] &  125.5 & 125.3 & 125.7 & 125.5 & 125.5  \\
$m_{h_{s}}$ [GeV] &  107.2 & 101.2 & 109.4 &105.6 & 95.4 \\
$m_{a_s}$ [GeV] &129.6 & 143.3 & 119.1 & 129.3 & 155.1  \\
$\left|\Delta_{\rm max}\right| $ & 3.7 & 3.7 & 7.8 & 7.8 & 5.2 \\
\hline
$m_{\widetilde{\chi}_1^0}$ [GeV] & 105.4 & 105.4 & 107.8 & 107.8 &  106.7 \\
$\Omega h^2$ &  0.12 & 0.12   & 0.12 & 0.12 & 0.12  \\
$\sigma_{\rm SI}$ [$10^{-45}$ cm$^2$] &  \hspace{.2cm} $1.26 $ \hspace{.2cm}  & \hspace{.2cm} 1.26 \hspace{0.2cm} & \hspace{0.2cm} 1.21 \hspace{.2cm} &  \hspace{.2cm} $1.21$  \hspace{.2cm} & \hspace{.2cm} $1.12 $  \hspace{.2cm}  \\
$\sigma_{\rm SD}$ [$10^{-42}$ cm$^2$] &  \hspace{.2cm} $5.12$  \hspace{.2cm} & \hspace{.2cm} 5.12 \hspace{0.2cm} & \hspace{.2cm} 11.61  \hspace{.2cm} &  \hspace{.2cm} $11.61$  \hspace{.2cm} &  \hspace{.2cm} 6.80  \hspace{.2cm} \\
$\langle \sigma v\rangle$ [$10^{-29}$ cm$^3/{\rm s}$] & 3.28 & 3.28 & 4.04 &  4.04 & $2.68$ \\

\end{tabular}
\caption{\label{tab:bm} \it\small The benchmarks considered in this study exemplifying the different phase transition possibilities in the NMSSM.  Aside from yielding various first-order phase transitions, parameters are chosen to yield a $\sim 125$ GeV Higgs with properties compatible with the resonance observed at the LHC, a viable neutralino dark matter candidate, and with the rest of the particle spectrum compatible with LHC searches and other constraints (see text).  The wino, gluino, and other sfermion soft breaking masses (besides $M_{\widetilde{t}}$) are set to $M_2=$ -600 GeV and $M_3=M_{sf}=1.5$ TeV for all benchmarks.  Note that the values of these masses do not significantly affect the scenarios we consider and can be increased if so desired. }
\end{table}

Before moving on to our analysis of the phase transition properties, in the remaining portion of this section we discuss the various phenomenological features of these points, explaining why they are good candidates for beyond-the-Standard Model physics and promising for electroweak baryogenesis.  The spectra are computed using \texttt{NMSSMTools 4.2.1} \cite{NMSSMTools}, with the DM properties calculated by \texttt{MicrOmegas 3.3} \cite{MicrOmegas}. We additionally check the Higgs sector using the \texttt{HiggsBounds 4.1} package \cite{Higgsbounds}.  We require all of our points to pass all relevant experimental constraints implemented in \texttt{NMSSMTools}, \texttt{MicrOmegas}, and \texttt{HiggsBounds}, except those arising from the muon $g-2$ and from requiring perturbativity up to the GUT scale.  The former can be ameliorated by e.g. including lighter sleptons and the latter by demanding that some new physics enter below the GUT scale \cite{oai:arXiv.org:hep-ph/0311349, oai:arXiv.org:1207.1435}.  Neither will affect any of our analysis of the phase transition or the particle phenomenology.   BM 1 and 2 have a Landau pole below $M_{GUT}$ as a result of the large values of $\lambda$ and small $\tan \beta$ considered in these cases.  We discuss the other relevant constraints in more detail below.

\subsection{Neutralino/Chargino Sector}
One of supersymmetry's virtues is that it can provide a viable dark matter candidate in the lightest supersymmetric particle (provided $R$-parity conservation), which in many cases is the lightest neutralino.  Consequently, we require all of our benchmarks to have a lightest neutralino LSP in agreement with both direct and indirect detection experiments and consistent with the observed relic density of dark matter \cite{WMAP9, PLANCK}
\begin{equation}
\label{eq:relden}
0.091\leq \Omega h^2\leq 0.138,
\end{equation}
where $h$ here is the local Hubble expansion parameter in units of 100 km/s/Mpc.  The interval quoted above corresponds to the $2\sigma$ limits from the WMAP 9-year data including $10\%$ theoretical uncertainty, and it also encompasses the range suggested by PLANCK data.  A viable neutralino DM candidate is not a requisite feature of SUSY; it is possible to instead have e.g. an axion \cite{axion} or gravitino \cite{gravitino} dark matter particle.  Requiring the LSP to make up 100\% of the observed dark matter density is not essential to our study and can be dropped in favor of another dark matter mechanism\footnote{Of course one must ensure that the LSP is not \emph{over}-abundant, regardless of the dark matter candidate.}.  However, one feature that we do find to be important for EWB is relatively low values of $\mu\lesssim 300$ GeV \cite{ Kozaczuk:2013fga}\footnote{Coincidentally, this range of $\mu$ is also favored by naturalness arguments \cite{Natural_1, Natural_2, Natural_3}.}.  This is because $\mu=\lambda v_s/\sqrt{2}$ and $v_s$ must generally be near the electroweak scale for the singlet to significantly impact the EWPT. For typical strongly first-order transitions involving the singlet (in the absence of large supercooling), $T_n \sim v_s(T_n)$, and so if $v_s\gg v$, one will typically find $v(T_n)/T_n \ll 1$ (for a one-step phase transition).  Low values of $\mu$ imply the existence of at least two relatively light neutralinos and one light chargino pair.  Higgsino-like neutralinos are not viable dark matter candidates, since they are always under-abundant.  Thus, requiring a neutralino LSP DM particle necessitates introducing either light bino or singlino states as the LSP.  

For the points we consider, the lightest neutralino is bino-like and close in mass to the singlino-like $\chi_2^0$, with $m_{\chi_2^0}-m_{\chi_1^0}$ approximately 200 to 2000 MeV.  This allows for efficient co-annihilation to suppress the thermal relic density, since pure bino dark matter is systematically over-abundant.  Consequently, our benchmarks all feature moderately light neutralinos and charginos, with the exception of the wino-like particles, which we take to have masses of 600 GeV so that their effects are largely decoupled from the standpoint of current experimental searches.  We did not attempt to choose masses consistent with gaugino mass unification, although it may be possible to do so.  Of course, there would be much more freedom in choosing these parameters if we were to drop the requirement of a viable neutralino dark matter candidate.  Note that, since we consider $\mu\lesssim 300$ GeV, all of our benchmarks will feature significant mass splitting between the chargino states, which suppresses the corresponding wino-Higgsino CP-violating sources relevant for electroweak baryogenesis \cite{Lee:2004we}.  However, there are several options for CP-violation which do not depend on the charginos.  Neutralinos can effectively source the baryon asymmetry \cite{Li:2008ez} ($M_1$ can be adjusted rather straightforwardly), and there are other potential sources in the NMSSM involving the scalars that have not been fully explored.  Thus, we believe that there is substantial room to include the necessary CP-violating sources required for successful EWB in the parameter space we consider.  We defer a more detailed study of CP-violating sources in the NMSSM to future study.

Since the LSP is bino-like, the cross-sections for spin-independent and spin-dependent scattering of $\chi_1^0$ with nucleons, denoted $\sigma_{\rm SI}$, $\sigma_{\rm SD}$, respectively, are in agreement with LUX \cite{LUX}, XENON100 \cite{XENON}, and  with other direct detection experiments for all the benchmark points we consider.  The zero-temperature annihilation rates for the various benchmark points are also compatible with limits from the Fermi large area telescope \cite{Fermi}.  Of course, dropping the requirement of a viable neutralino dark matter candidate, the signals predicted for direct- and indirect-detection experiments will be significantly weakened, since the LSP will only make up a fraction of the dark matter abundance.

This set-up is also comfortably compatible with current LHC limits on electroweak-ino production.  While the winos are heavy, the Higgsino-like charginos and neutralinos in all of our benchmarks have masses between 200--300 GeV, and decay with almost 100$\%$ branching ratio to final states involving $\chi_2^0$. Since the mass splitting between $\chi_1^0$ and $\chi_2^0$ is small, we have typically $BR(\chi_2^0\rightarrow \chi_1^0 \gamma)\approx 100\%$.  The resulting photon is very soft and although the decay can result in a displaced vertex, the decay products will simply be counted as missing energy, since such soft photons fall well below the trigger thresholds in the ATLAS and CMS electromagnetic calorimeters.  Thus, from the standpoint of LHC searches, the parameter space we explore can be thought of as an effective light singlino-Higgsino scenario, as considered in e.g. Refs.~\cite{Ellwanger:2013rsa, Kim:2014noa}.  The most relevant LHC constraints are those arising from ATLAS \cite{ATLAS_EW} and CMS \cite{CMS_EW} searches for trileptons with missing transverse energy.  A simple application of the limits from Ref.~\cite{Ellwanger:2013rsa} shows that Benchmarks 1--5 lie well within the allowed region of parameter space, although they may be probed at the 14 TeV LHC \cite{Kim:2014noa}.  As a check, we have simulated events using the \texttt{Madgraph} \cite{Madgraph}-\texttt{Pythia}\cite{Pythia}--\texttt{Delphes}\cite{Delphes} pipeline, utilizing the program \texttt{CheckMATE} \cite{CheckMATE} to perform the cuts and check against existing ATLAS searches.  As expected, the benchmarks satisfy current constraints from trilepton, mono-jet \cite{ATLAS_monojet}, jets+ MET \cite{ATLAS_0lepton}, and di-lepton \cite{ATLAS_dilepton} searches at ATLAS (the corresponding CMS searches have not yet been validated for use with \texttt{CheckMATE}).  While the collider phenomenology of the regions we consider here deserves a more detailed study, it is beyond the scope of this work.  Here, we simply emphasize that the benchmark points listed in Table~\ref{tab:bm} all satisfy current LHC constraints on chargino and neutralino searches.

\subsection{Sfermion Sector}
Heavy superpartners are not fundamentally required from the standpoint of electroweak baryogenesis, but rather by their non-observation so far at the LHC.  For $m_{\chi_1^0}\sim 100$ GeV, as in our benchmarks, ATLAS and CMS currently exclude stops with masses less than about 700 GeV \cite{ATLAS_stop1, ATLAS_stop2, CMS_stop1}, albeit with some caveats and potential loopholes.  Nevertheless, we take the stop masses to be at or above 1 TeV.  All other sfermions are also assumed to have larger masses, set to 1.5 TeV for all benchmarks.  Although stop masses above a TeV are already  in some tension with naturalness \cite{Natural_1, Natural_2, Natural_3}, all of our benchmarks fall within the $10-30\%$ fine-tuning range as computed in \texttt{NMSSMTools 4.2.1} \cite{NMSSMTools, oai:arXiv.org:1107.2472} and are arguably quite natural in this sense.  To quantify the amount of tuning for each of our benchmarks, in Table~\ref{tab:bm} we show the value of $\Delta_{\rm max}$ for each point, defined by \cite{oai:arXiv.org:1107.2472}
\begin{equation}
\Delta_{\rm max} \equiv \operatorname{Max} \left\{\Delta_i^{\rm GUT}\right\}, \hspace{.3 cm} \Delta_i^{\rm GUT} =\left| \frac{\partial \log (M_Z)}{\partial \log (p_i^{\rm GUT})}\right|
\end{equation}
where $p^{\rm GUT}_i$ are the input parameters of the theory at the GUT scale.  The amount of fine-tuning is given approximately by $1/\Delta_{\rm max}$.  For all benchmarks, $p^{\rm GUT}_{\rm max}=m_{h_u}^2$, except for BM 2, which features $p^{\rm GUT}_{\rm max}=A_{\lambda}$.  

For the sfermion mixing parameters, we choose all tri-scalar couplings to be less than 2 TeV.  This, again, takes its cue from naturalness, but also avoids vacuum stability issues that can arise for large $A_t$ \cite{Chowdhury:2013dka, Blinov:2013fta}.  From the standpoint of the electroweak phase transition in this case, the precise values of the sfermion masses and mixing parameters (other than those for the stops) are immaterial: their effects are decoupled since they are far above the EW scale and because of their small couplings to the Higgs sector.  We choose specific values only to concretely demonstrate the phenomenology of the various benchmark points.  

The gluino mass is set to 1.5 TeV for all benchmarks.  This choice is again motivated by the lack of evidence for a light gluino at the LHC \cite{ATLAS_gluino, CMS_gluino} and by naturalness considerations.  The precise value of the gluino mass has little impact on the phenomenology we consider here.

\subsection{Higgs Sector}
Of course all of our benchmarks should also be consistent with the Higgs discovery, i.e. feature a Higgs boson in the range $124$ GeV $\lesssim m_{h,SM}\lesssim 127$ GeV \cite{ATLAS_Higgs, CMS_Higgs} with couplings very similar to those predicted by the Standard Model \cite{ATLAS_couplings, CMS_couplings}.  More precisely, we require that all of our benchmarks predict signal strengths that match those observed by the ATLAS and CMS collaborations in the various production and decay channels.  To this end, one usually defines the signal strength parameter $\mu(X,Y)$ as
\begin{equation} \label{eq:muxy}
\mu(X,Y)\equiv \frac{\sigma_{h,X}\cdot BR\left(h\rightarrow Y\right)}{\sigma_{h_{SM},X}\cdot BR\left(h_{SM}\rightarrow Y\right)}
\end{equation}
where $\sigma_{h,X}$ is the total production cross-section of $h$ via the process $X$, where $X=$ ggF (gluon-gluon fusion), ttH (associated production with a top quark pair), VBF (vector boson fusion), or VH (associated production with a gauge boson).  Here $h$ is taken to be the SM-like Higgs in the NMSSM (the second-lightest CP-even Higgs for our benchmarks), and $h_{SM}$ is the Standard Model Higgs boson.  $BR(h\rightarrow Y)$ is the branching ratio of $h$ to final state $Y$ where $Y=\gamma \gamma$, $VV$, $b\bar{b}$, or $\tau \tau$, and similarly for $h_{SM}$.  Typically, the production modes are grouped and analyzed together as $X_1\equiv$ ggF+ttH and $X_2\equiv$ VBF+VH (see e.g. Ref.~\cite{Belanger:2013xza} for further discussion).  Measurements from ATLAS and CMS, together with those from the Tevatron, can then be used to derive constant likelihood contours in the $\mu(X_1,Y)-\mu(X_2,Y)$ planes by means of a global fit.  This has been done recently in Refs.~\cite{Giardino:2013bma, Belanger:2013xza}.

\begin{figure*}[!t]
\mbox{\includegraphics[width=0.435\textwidth,clip]{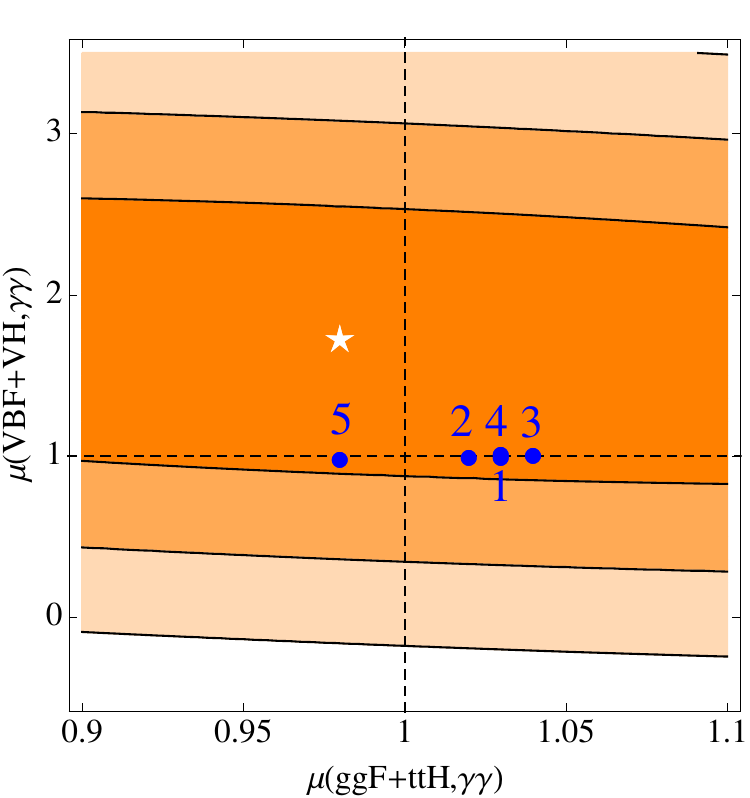}\quad \includegraphics[width=0.45\textwidth,clip]{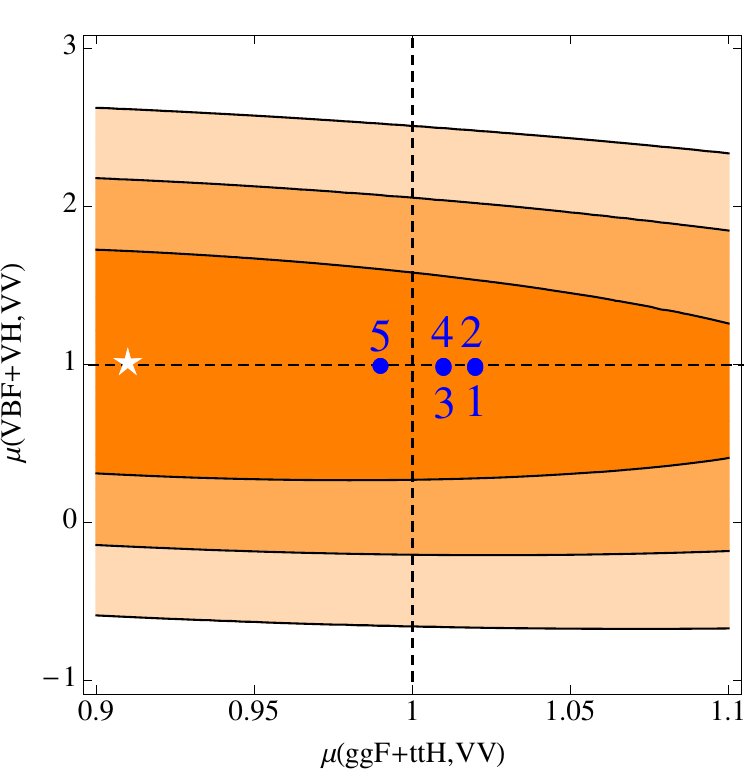}}\caption{\label{fig:signal} \it \small Signal strengths for the various Higgs prodution and decay channels for our benchmark points (labeled 1--5), compared with the global fit in Ref.~\cite{Belanger:2013xza} obtained using current ATLAS, CMS, and Tevatron data.  On the left we consider the diphoton rate arising from vector boson fusion (VBF) + associated production with a gauge boson (VH), and from gluon gluon fusion (ggF) + associated production with a top quark pair (ttH). On the right we plot the corresponding results for vector boson final states.  The white star indicates the current best-fit point from Ref.~\cite{Belanger:2013xza}, while the shaded areas correspond to 68\%, 95\%, and 99.7\% C.L. regions from darkest to lightest, respectively.  All the benchmark points lie within the 68\% CL regions for the observed signal strengths.  The $b\bar{b}/\tau\tau$ ellipses are not shown, since all the benchmarks lie very close to the best fit point in this plane.  All of our benchmark points feature a very Standard Model-like Higgs in good agreement with observation.}
\end{figure*}

To check whether or not our points agree with experimental data on the Higgs signal rates, we compute $\mu(X,Y)$ for all relevant channels using \texttt{NMSSMTools 4.2.1} \cite{NMSSMTools} and verify that they lie within the 95\% C.L. regions derived in Ref.~\cite{Belanger:2013xza}.  The results for the various benchmarks are shown in Fig.~\ref{fig:signal}, superposed on the likelihood contours obtained from the $\chi^2$ in Ref.~\cite{Belanger:2013xza}.  The figure shows good agreement between our points and the results from ATLAS, CMS, and the Tevatron. All benchmarks lie within the 68\% C.L. regions and are very Standard Model-like; if the enhanced $h\rightarrow \gamma\gamma$ rate continues to decrease in statistical significance, all points will move into even better agreement with the data.  We have also cross-checked all of our benchmarks with \texttt{HiggsSignals 1.1}\cite{HiggsSignals}, which takes into account the various correlations and systematics that enter the fit.  This can be important in cases when there are multiple Higgs bosons near 125 GeV, as is the case for several of our benchmarks.  We again find good agreement with current observation for all points considered. 

We must also ensure that the rest of the Higgs sector does not violate the current constraints from LEP, the Tevatron, or the LHC.  This is precisely what is checked by \texttt{HiggsBounds}.  In all of our benchmarks, the scalar closest in mass to the SM-like Higgs is singlet-like, with couplings of order 10\% or less of those for a SM Higgs boson with the same mass.  These suppressed couplings make it difficult to detect these states and allow them to be even lighter than the SM-like Higgs, as will be the case for all the points we consider.  In fact, in the NMSSM, scalars and pseudoscalars can be extremely light and still compatible with current collider and meson decay limits, provided the mixing is small \cite{Kozaczuk:2013spa} (see e.g. Ref.~\cite{Buckley:2014ika} for a more detailed discussion of possible strategies to search for these additional states at the 14 TeV LHC). Additionally, for all of the points considered, $\tan \beta$ is low and the MSSM parameter $m_A$, which sets the mass of the charged, CP-odd, and third neutral Higgs bosons, is between 370--410 GeV, making searches for charged Higgs states difficult as well.  For all the cases we consider, the strongest bounds on the Higgs sector come from the couplings of the SM-like Higgs and all other Higgs constraints are satisfied by a comfortable margin.  

\section{The Phase Transitions and their Properties}\label{sec:EWPT}

With the relevant particle phenomenology in hand, we now turn to analyzing the phase transition properties for the various benchmarks described above.  First, we define and discuss the parameters of interest from the standpoint of electroweak baryogenesis and cosmology, and detail how we compute them.  We then move on to present our results for the various points considered in Sec.~\ref{sec:results}.

\subsection{Studying Phase Transitions in the NMSSM}\label{sec:PT}
 
The cosmological phase transitions predicted by the NMSSM are of interest primarily because they may be able to support successful electroweak baryogenesis.  If a first-order phase transition is strong enough, it can provide the out-of-equilibrium dynamics necessary to quench the processes which wash out the baryon asymmetry after it is produced, namely the $SU(2)$ sphalerons.  The requirement that the $SU(2)$ sphaleron rate be sufficiently suppressed inside the bubble is usually phrased in terms of the order parameter $v(T_n)/T_n$, where $T_n$ is the nucleation temperature of the bubble (defined in more detail below).  Aside from the gauge-dependence inherent in this quantity (discussed in Sec.~\ref{sec:NMSSM_Veff}), the correct baryon number preservation condition depends on how large the baryon washout can be.  This in turn depends on the strength of the CP-violating sources generating the chiral current, as well as the details of the diffusion of the various charge densities in front of the bubble wall \cite{Patel:2011th}.  For example, as a result of these uncertainties, the correct baryon number preservation condition in the Standard Model can range from $v(T_n)/T_n \gtrsim 0.4$--$1.4$, depending on the details of the CP-violation and transport \cite{Patel:2011th}.  

Following the usual convention, we will define a ``strongly first-order phase transition" as a first-order transition such that 
\begin{equation}
\frac{\Delta \phi}{T_n}\geq 1
 \end{equation}
Here, we have defined the slightly more general quantity $\Delta\phi=\sqrt{\sum (\phi_{i,n}-\phi_{i,0})^2}$, where $\phi_{i,n}$ is the value of the field $i$ in the low-temperature minimum, and $\phi_{i,0}$ corresponds to the field value in the high-temperature phase before nucleation.  The sum runs over the fields of interest.  For the NMSSM case, we will be primarily interested in $\Delta\phi_{SU(2)}$ (corresponding to $\phi_i = h_{u,d}$) and $\Delta\phi_s$ (corresponding to $\phi_i=s$).  Points with $\Delta\phi_{SU(2)}\geq1$ and $h_{u,0}=h_{d,0}=0$ may lead to successful electroweak baryogenesis.  If either $h_{u,0}\neq0$ or $h_{d,0}\neq0$, the sphalerons will already be suppressed in the space-time region outside the bubble, and so the contribution to the baryon asymmetry from electroweak baryogenesis will be suppressed.  Still, such a situation may be of interest from the standpoint of cosmological signatures such as gravitational radiation, or extended electroweak baryogenesis scenarios.  For the same reason, we will also consider transitions with $\Delta \phi_{SU(2)}=0$ but $\Delta \phi_s\geq 1$.  

For each of our benchmark points, we therefore calculate the high- and low-temperature VEVs and the difference between these VEVs in both the $SU(2)$ direction ($\Delta\phi_{SU(2)}$) and the singlet direction ($\Delta \phi_s$).  In addition to the phase transition order parameter, we will also compute the change in energy density ($\Delta \rho$) at the transition, and  the change in pressure ($\Delta p$).  
Since the finite-temperature effective potential $V(\phi, T)$ is equal to the free energy density, the change in total energy density is given by
\begin{align}
\Delta \rho = \sqof{V\of{\phi_0, T} - {dV\of{\phi_0,T}\over dT} \cdot T}_{T=T_n} - \sqof{  V\of{\phi_n,T}- {dV\of{\phi_n,T}\over dT} \cdot T }_{T=T_n}
\end{align}
where $\phi_0$ and $\phi_n$ are the (three-dimensional) field values in the false and true vacua, respectively. 
Note that if there is no supercooling, $\Delta\rho$ is the same as the transition's latent heat.
The quantity $\Delta\rho$ provides another measure of the strength of the transition, with larger $\Delta\rho$ corresponding to more strongly first-order transitions.
The pressure difference between the phases is simply the change in the finite-temperature effective potential from the high- to low-temperature VEV:
\begin{align}
\Delta p = V\of{\phi_0,T_n} - V\of{\phi_n,T_n}.
\end{align}

Also of crucial importance from the standpoint of electroweak baryogenesis (and any other microphysical calculation involving the bubble wall) are the details of the bubble wall profile, $\phi(z)$ (here $z$ is a spatial coordinate in the frame comoving with the bubble wall and $\phi$ is a vector in field space).  Several of the $CP$-violating sources which enter the microphysical calculations of the baryon asymmetry are proportional to $d\beta/dt$ \cite{Lee:2004we}, and the variation of the Higgs VEVs in the bubble wall is crucial for obtaining non-vanishing CP-violating sources for the chiral current.  In the literature, $d\beta/dt$ is typically estimated as \cite{Lee:2004we}
\begin{equation}
d\beta/dt\simeq \Delta \beta v_w/L_w.
\end{equation}
Here $\Delta \beta$ is defined as the change in the angle $\beta = \arctan(h_u/h_d)$ from the high-temperature VEV to the low-temperature VEV, $L_w$ is the instanton bubble wall widths in the $SU(2)$ direction, and $v_w$ is the velocity of the bubble wall in the frame of the plasma far away from the wall outside the bubble.  We will thus calculate $\Delta \beta$, $L_w$, and $v_w$ for our various benchmark points, as well as $L_s$, the wall width in the singlet direction, as this quantity will enter the CP-violating sources involving the singlet VEVs.

While much previous work relied on various ansatz{\"e} for the wall profile, we solve for the tunneling solution numerically.  We must therefore clearly define what we mean by the parameters $L_w$ and $L_s$.  In practice, the widths are defined as the distance in $r$ space over which the relevant field $\phi$ drops from 73\% to 27\% of its total height $\Delta\phi$.  This convention is chosen to coincide as closely as possible to the oft-used definition of $L_{w}$ as the parameter entering the hyperbolic tangent profile $\phi(x)=\Delta\phi/2(1+\tanh x/L_{w})$. Our definition of $\Delta\beta$ corresponds to the difference in $\beta$ between the spacetime points where the field value $\phi$ comes within 5\% of its value in the minima, i.e.
\begin{equation}
\Delta\beta\equiv \beta(x_{\rm low})-\beta(x_{\rm high})
\end{equation}
where $x_{\rm low}$, $x_{\rm high}$ are defined via
\begin{equation}
\begin{aligned}
\phi(x_{\rm low})&=\phi_0+0.95\Delta\phi\\
\phi(x_{\rm high})&=\phi_0+0.05\Delta\phi
\end{aligned}
\end{equation}
and $\phi=\sqrt{h_u^2+h_d^2}$.

Strong, singlet-driven transitions can give rise to observable gravity wave signatures \cite{Ashoorioon:2009nf}.  Thus, in addition to the bubble wall properties, we also compute the spectrum of gravitational radiation produced by the strongest first-order transitions. For this we require the relative change in energy density over the transition, $\alpha$, and the inverse of the duration of the transition\footnote{In most previous studies, this quantity was denoted as $\beta$.}, $\zeta$. From Refs.~\cite{Wainwright:2011qy,Huber:2008}, we have
\begin{align}
\alpha ={ \Delta \rho \over \rho_\mathrm{rad}},
\end{align}
where  $\rho_\mathrm{rad} = {g^* \pi^2 \over 30} T_n^4$
and $g^*$ is the number of relativistic degrees of freedom, taken here to be $100$.  Meanwhile, 
\begin{align}
{\zeta \over H} = \sqof{T \cdot {d (S_3/T) \over dT}}_{T=T_n}.
\end{align}
where
$S_3$ is the Euclidean action, and $H$ is the Hubble expansion rate during the transition.  These quantities allow us to calculate the gravity wave overall amplitude
\begin{align}
h^2 \tilde{\Omega}_\mathrm{GW} = 1.67 \times 10^{-5} \tilde{\Delta} K^2 \ofp{  H\over \zeta}^2 \left({ \alpha \over \alpha + 1}\right)^2 \ofp{  100\over g^* }^{1/3},
\end{align}
and peak frequency
\begin{align}
\tilde{f} = 16.5\times10^{-3} {\rm mHz} \ofp{   \tilde{f}_* \over \zeta  } \ofp{  \zeta \over H}  \ofp{   T \over 100 \hspace{.1cm}{\rm GeV} }  \ofp{   g^* \over 100}^{1/6}
\end{align}
where
\begin{align}
\tilde{\Delta}\of{v_w} = { 0.11 v_w^3 \over 0.42 + v_w^2 }\\
\ofp{   \tilde{f}_* \over \zeta }\of{ v_w}  = {    0.62 \over 1.8 - 0.1 v_w + v_w^2  }
\end{align}
and $K$ is an efficiency factor, given\footnote{In Ref.~\cite{Huber:2008}, this quantity is denoted as $\kappa$.} as a function of $\alpha$ in e.g. Ref.~\cite{Huber:2008}.

We employ the {\tt CosmoTransitions} package~\cite{Wainwright:2011kj} to study the phase structure and thermal tunneling properties of each of the benchmark points. The full one-loop finite-temperature effective potential (Eq.~\ref{eq:2HDS_potential}) is directly input into the program along with the zero-temperature electroweak VEV. The minimum at the VEV is traced upwards in temperature (the VEVs are generally temperature dependent) until it either disappears or merges into a distinct high-temperature phase via a second-order transition\footnote{The \texttt{CosmoTransitions} package designates a transition as second-order if the associated high-temperature and low-temperature phases have non-overlapping temperature domains, the gap between the two domains is smaller than the resolution of phase tracing routine (typically on the order of 0.1 GeV), and the two phases are proximate at the transition temperature. The end-point of a phase is given by the temperature at which the Hessian matrix has an approximately zero eigenvalue. Therefore, `second-order' is a numerical rather than analytic designation, and very weakly first-order transitions may get misclassified as second-order. However, from the viewpoints of electroweak baryogenesis and gravitational radiation production, second-order and very weakly first-order transitions are functionally equivalent.}. Either way, a new minimum is found at the temperature at which the first phase ends, and the structure of the new phase can likewise be found. In this way, the {\tt CosmoTransitions} package determines the theory's complete temperature-dependent phase structure. In all of our benchmarks, the low-temperature phase is just the electroweak phase determined by the electroweak VEV, and the high-temperature phase resides at the origin of field space. However, different benchmarks exhibit qualitatively different intermediate phases. They can generally lie along either the singlet or $h_u$ directions, a mixture of several directions, or be missing entirely (that is, the system can transition directly from the high-$T$ to electroweak phase). We detail the features of individual benchmarks in Sec.~\ref{sec:results} below.

Once the phase structure is found, it is straightforward to calculate the critical temperatures $T_c$ at which any two temperatures have degenerate minima (corresponding to equal free energy and equal pressure). The {\tt CosmoTransitions} package then finds the nucleation temperatures $T_n$ below $T_c$ such that there is unit probability to nucleate one bubble of lower-energy phase within a higher-energy background phase per horizon volume per Hubble time. Numerically, to determine the nucleation temperature we use the rough criteria~\cite{QuirosReview} that $S_3(T_n)/T_n \sim 140$, where the bubble action $S_3$ represents the energy of a critical bubble whose surface tension exactly balances the pressure gradient across its wall. 

The profile of each critical bubble is found using the \texttt{CosmoTransitions pathDeformation} module. First, the path that a bubble takes through field space is assigned a tentative initial value. A one-dimensional version of the equations of motion which govern the bubble profile is solved along the initial path using an overshoot / undershoot method. The path is then slightly deformed to reduce the magnitude of perpendicular forces, after which the one-dimensional equations are solved again. By iterative deformations, the \texttt{pathDeformation} routines converge towards the correct bubble profile. We check our solutions by choosing different initial paths and verifying that they produce the same result. In particular, an initially straight path and a path that initially crosses the saddle point separating two phases will both approach the correct solution, but from different directions.

If the critical bubble is very thick walled, then its center may be significantly displaced from the low-temperature VEV. Once it starts growing, its center will quickly roll down to the bottom of the minimum. Therefore, the wall profile of an expanding bubble may not match that of the incipient critical bubble. To account for this possible discrepancy, we calculate the quantities $L_w$, $L_s$, $v_w$ and $\Delta\beta$ for bubble walls moving at constant velocity with constant friction; we dub these ``late time" bubble profile parameters for this reason. Details of this calculation are given in Sec.~\ref{sec:vw}, below.


\subsection{Checking for Runaway Walls} \label{sec:runaway}

Since we are interested in the possibility of electroweak baryogenesis at the various ($SU(2)$) phase transitions, one additional crucial criterion to check is whether or not the electroweak bubble wall can run away, i.e. $v_w\rightarrow 1$.  This can occur if the pressure differential driving the acceleration of the wall cannot be balanced by the drag exerted on it by the plasma.  Runway walls are detrimental to electroweak baryogenesis: if $v_w$ is larger than the sound speed of the plasma, $v_s\simeq 1/\sqrt{3}$, sphaleron processes acting on the diffusing chiral currents ahead of the wall are too inefficient to source a sizable baryon asymmetry.  Successful transport-driven electroweak baryogenesis thus fundamentally requires sub-luminal, in fact sub-sonic, wall velocities.

An important first test for wall runaway has been suggested in Ref.~\cite{runaway}.  Physically, there are effectively two competing forces acting on the bubble wall: the vacuum free-energy density difference between the phases, $ V(\phi_0,T=0)-V(\phi_n, T=0)\equiv\Delta V(T=0)$, acting as a driving force, and the pressure exerted by the plasma particles on the scalar field background, $F_p/A$ ($A$ here is the area of the wall, neglecting curvature). A necessary, but not sufficient \cite{Konstandin:2010dm, Megevand:2013hwa}, condition for a runaway wall is that $F_p/A < F_{vac}/A = \Delta V(T=0)$.  The pressure exerted by the plasma in the relativistic limit is given by \cite{runaway, Espinosa:2010hh}
\begin{equation}
\frac{F_p}{A} = \sum_i \left|n_i\right|\left(m_i^2 (\phi_0)-m_i^2(\phi_n)\right) \int \frac{d^3 p}{(2\pi)^3 2 E_{i,p}(\phi_0)} f_{i,p}^{eq}(\phi_0)
\end{equation}
where $n_i$ is the number of degrees of freedom for species $i$, $m_i(\phi)$ are the field-dependent masses (including the thermal masses for the bosons), $E_{i,p}(\phi)=\sqrt{p^2+m_i^2(\phi)}$, and where $f_{i,p}^{eq}$ are the \emph{equilibrium} distribution functions.  Since the wall's motion is assumed to be ultra-relativistic, the passage of the wall changes the masses sharply but leaves the distribution functions as they were in the symmetric phase to leading order in $1/\gamma$ (where $\gamma=1/\sqrt{1-v^2}$) \cite{Espinosa:2010hh}.

As pointed out in Ref.~\cite{runaway}, the above expression is equivalent to the free-energy density difference between the minima in the so-called \emph{mean field} $T\neq 0$ thermal effective potential, $\widetilde{V}_T$ \cite{runaway, Espinosa:2010hh}.  $\widetilde{V}_T$ is simply given by a Taylor expansion of $V_{1,T}$ around the symmetric minimum in field space, truncated at quadratic order in the field-dependent masses, i.e. \cite{runaway, Espinosa:2010hh}
\begin{equation}
\widetilde{V}_T (\phi, T \neq0) \equiv V_{T} (\phi_0, T\neq0) + \sum_{i} \left(m_i^2(\phi)-m_i^2(\phi_0)\right) \frac{dV_T(\phi_0,T\neq0)}{dm_i^2}.
\end{equation}
With this definition, the condition that must be satisfied for the wall to run away can be re-phrased as $\widetilde{V}_T(\phi_n, T_n)-\widetilde{V}_T (\phi_0,T_n) < V_1(\phi_0,T=0)-V_1(\phi_n,T=0)$.  Re-arranging, and defining the full 1-loop mean field effective potential $\widetilde{V}(\phi,T)\equiv V_0(\phi)+V_1(\phi,T=0)+\widetilde{V}_T(\phi,T)$, the runaway condition becomes
\begin{equation} 
\widetilde{V}(\phi_0,T_n)-\widetilde{V}(\phi_n, T_n) > 0,\ \ \  \leftrightarrow\ \ \  {\rm Runaway \hspace{0.2cm} solution \hspace{0.2 cm} exists.}
\end{equation}
This suggests the following criterion for determining whether the wall is safe from runaway \cite{runaway}: if tunneling to the broken minimum $\phi_n$ is not energetically favored in the mean field potential, the wall cannot run away.  Thus, for all of our benchmarks, it suffices to compute both the full effective potential and the mean field potential along the tunneling direction; if the symmetry-breaking minimum disappears or is raised above the symmetric minimum in the mean field limit, the wall will remain sub-luminal.  We will check against this criterion for all of the points we consider.

\subsection{Results}\label{sec:results}

\begin{table}
\begin{tabular}{ c | c  | c | c | c | c | c }
										&	 BM 1			&	BM 2				&	{BM 3}		&	\multicolumn{2}{c|}{BM 4	}	 				& BM 5 					\\
\hline	
Direction									&	s+h				&	s+h				&	s			&		s				&	h				&	s+h					\\
\hline
$\Delta \phi_{SU(2)}/T_n$		&	1.3				&	2.0				&	0			&		0				&	0.42			&	2.0				\\
$\Delta \phi_\mathrm{s}/ T_n$		&	2.6			&	3.6				&1.1			&		1.6			&	0.20			&	4.2				\\
\hline
$T_c$ (GeV)		&		153.4		&	142.0	&		188.7		& 170.9		& 	165.4		&		141.2 \\
$T_n$ (GeV)		&		140.1		&	108.4			&186.5			&		165.7			&	165.4			&	103.0					\\
\hline
high-T VEV						&	(0,0,0)			&	(0,0,0)			&	(0,0,0)		&		(0,0,0)			&	(0,0,260)			&	(0,20,0)					\\
low-T VEV							&\hspace{.08cm} 	(96,150,362)\hspace{.08cm} 	&\hspace{.08cm} 	(121,183,390)\hspace{.08cm} 	&\hspace{.08cm} 	(0,0,199)	\hspace{.08cm} 	&\hspace{.08cm} 		(0,0,258)	\hspace{.08cm} 		&\hspace{.08cm} 	(29,62,292)\hspace{.08cm} 	&	\hspace{.08cm} (118,191,430)		\\
\hline
$\Delta p$	 ($\mathrm{GeV}^4$)	
							&	$(79.4)^4$		&	$(100.4)^4$		&$(34.4)^4$		&		$(46.4)^4$		&	$(14.4)^4$		&	$(104.9)^4$			\\
$\Delta \rho$ ($\mathrm{GeV}^4$)	
							&	$ (151.2)^4$		&	$(144.5)^4$		&$(110.2)^4$		&		$(116.7)^4$		&	$(100.1)^4$		&	$(143.4)^4$			\\
\hline
$\left|\Delta \beta\right|$					&	$0.32$			&	N/A				&	N/A		&		N/A				&	$0.029$			&	 N/A			\\
$L_w$ $T_n$	&	$4.0$			&	N/A				&	N/A	&		N/A			&	$23.7$			&	N/A				\\	
$L_s$ $T_n$ 			&	$5.9$			&	N/A				&	$16.4$	&		$11.9$			&	$20.5$			&	N/A				\\
%
%
\end{tabular}
\caption{\it \small Properties of the first-order transitions of BM 1--5.  Here, ``Direction'' indicates the field direction in which the transition occurs. ``s'' stands for the singlet direction, and ``h'' stands for the $SU(2)$ direction. $\Delta \phi_{SU(2)}$ is defined as the change in field values in just the SU(2) directions: $ \Delta \phi_{SU(2)} \equiv \sqrt{ \Delta h_u^2 + \Delta h_d^2 }$, and $\Delta \phi_s$ is the change in the singlet field value: $\Delta \phi_s \equiv \Delta s$ .  The three-dimensional field $\phi\equiv(h_d, h_u, s)$ transitions from the ``high-T VEV'' to the ``low-T VEV'' at nucleation temperature $T_n$. $\Delta \beta$ is the change in $\beta$ from the high-temperature VEV to the low-temperature VEV, where $ \beta  \equiv \arctan\left({ \langle h_u \rangle / \langle h_d \rangle} \right)$.  Note that $\Delta\beta$ is only defined for transitions in the SU(2) directions, so is not listed for singlet-only transitions. The late-time bubble wall profile parameters are only calculated for walls moving with constant velocity and friction, and so are not listed for benchmarks with runaway solutions.
}
\label{tab:outtab}
\end{table}

\begin{figure}[!t]
\includegraphics[width=\textwidth]{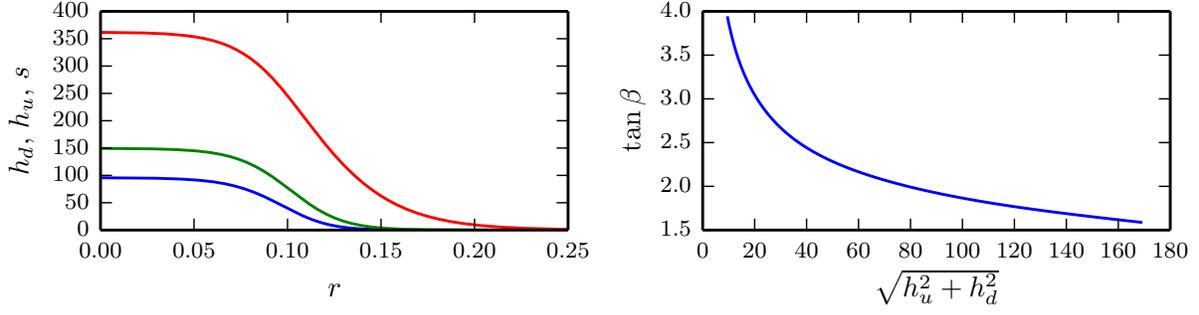}
\caption{
\it \small Late time bubble wall profiles (left) and $\tan \beta$ (right) for the strongly first order electroweak phase transition of BM 1.
}
\label{fig:tanbeta_and_walls}
\end{figure}

\begin{figure}[!t]
\includegraphics[width=\textwidth]{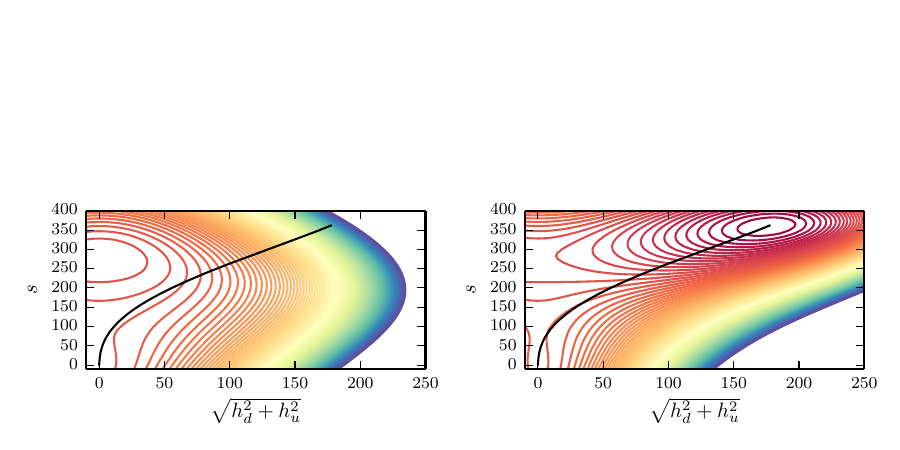}
\caption{
\it \small $V$ plotted against $s$ and $\sqrt{h_d^2+h_u^2}$ with $\tan\beta$ fixed. On the left, $\tan \beta = 4.01$, which is its value just outside the bubble wall, \textit{i.e.} where $\phi = \phi_\mathrm{low} + 0.95\Delta\phi$. On the right $\tan \beta = 1.59$, its value just inside the bubble wall (where $\phi = \phi_\mathrm{low} + 0.05\Delta\phi$), where indeed the potential minimum is at a nonzero value of $s$ and $\sqrt{h_d^2+h_u^2}$. The black lines are the late time tunneling paths.
}
\label{fig:VContours}
\end{figure}

\begin{figure*}[!t] 
\subfigure{\includegraphics[width=0.45\textwidth]{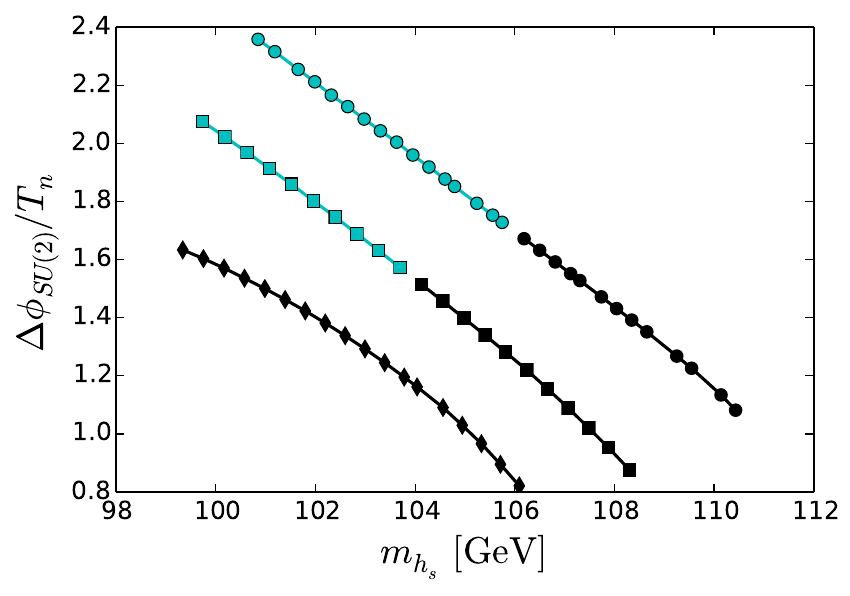}}
\subfigure{\includegraphics[width=0.45\textwidth]{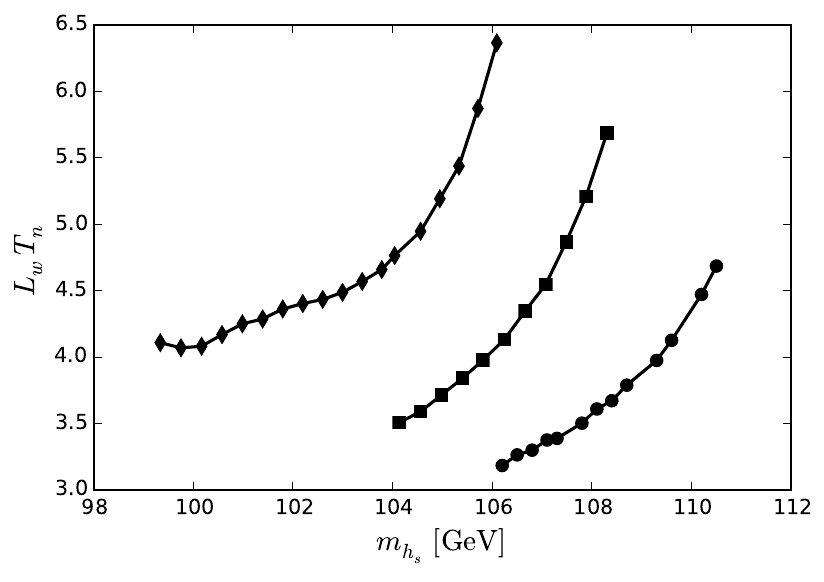}}
\subfigure{\includegraphics[width=0.45\textwidth]{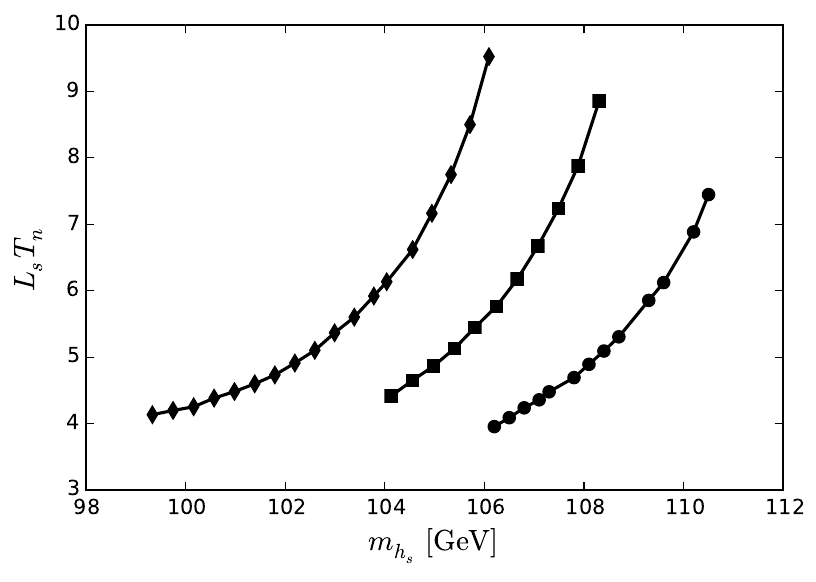}}
\subfigure{\includegraphics[width=0.45\textwidth]{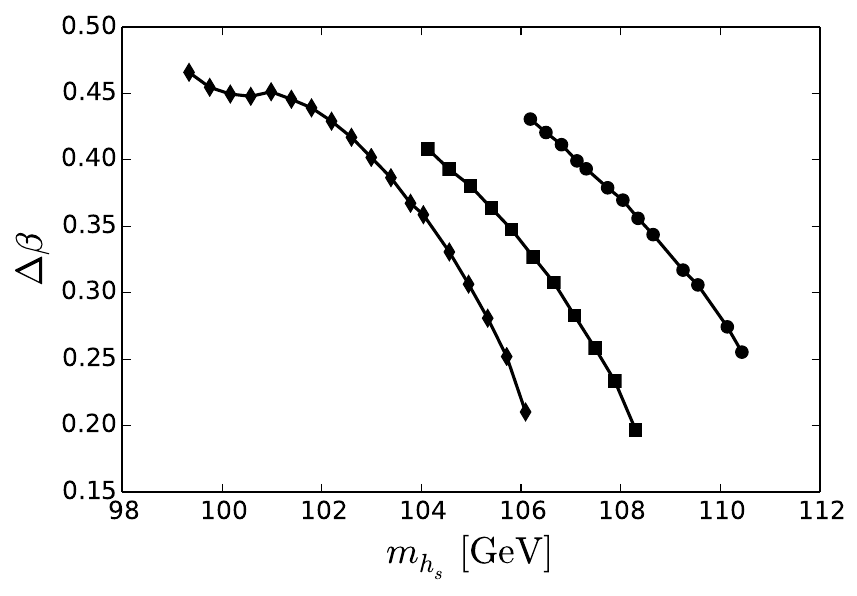}}
\caption{ 
\label{fig:results}
\it \small Results for strongly first order one-step electroweak phase transitions at different values of $m_{h_s}$ for Sets I (circles), II (diamonds), and III (squares).  Shown are the EWPT order parameter, $SU(2)$ wall width, singlet wall width, and $\Delta \beta$, which are quantities relevant for investigations of electroweak baryogenesis.  The singlet-like Higgs mass is varied by varying $A_{\kappa}$ as described in the text with all other parameters fixed.  The rest of the spectrum varies very little across the scanned points, with the phenomenology as presented in Table~\ref{tab:bm}.  Black points have bubble walls that are guaranteed to be sub-luminal, while the cyan points admit a runaway solution.  Note that the late-time bubble wall profile parameBters are only calculated for walls moving with constant velocity and friction, and so are not shown for points with runaway solutions.
}
\end{figure*}
We can now move on to our main results, summarized in Table~\ref{tab:outtab}.  Consider first the case of a one-step electroweak phase transition (EWPT).  This is exemplified by BM 1, which features a phase transition at $T_n=140.1$ GeV.  The field transitions from the origin to a phase in which all three field directions have non-zero VEVs.  This transition is strongly first-order in both the $SU(2)$ and singlet directions, and proceeds after a relatively small amount of supercooling ($T_c=153.4$ GeV).  This point does not feature a runaway bubble wall solution, and thus can potentially lead to successful electroweak baryogenesis.  To see this, we plot the mean field and full effective potential in the upper left panel of Fig.~\ref{fig:runaway}.  Clearly, the broken minimum is raised above the origin in the mean field limit, and so, by the results of Sec.~\ref{sec:runaway}, the bubble wall must approach a stationary state.

We show the details of the bubble wall profile and tunneling path corresponding to this point in Figs.~\ref{fig:tanbeta_and_walls} and~\ref{fig:VContours}.  The profile is characterized by $L_w=4.0/T_n$, $L_s=5.9/T_n$, and $\left|\Delta\beta\right|=0.32$.  Note the large value of $|\Delta\beta|$ as compared to the MSSM case \cite{Moreno:1998bq}.  This is a very attractive feature from the standpoint of electroweak baryogenesis, since large values can allow for smaller CP-violating phases in the sources, resulting in less stringent bounds from electric dipole moment experiments.  Although at tree-level (at high energies) the singlet couples in the same way to $h_u$ and $h_d$, after integrating out the stops and evolving the parameters down to the electroweak scale, this is no longer the case.  Note that smaller values of $\Delta \beta$ were found previously for the general NMSSM in Ref.~\cite{Huber:2000mg} (on the order of $10^{-3}$, close to MSSM values). However, we believe this discrepancy can be explained by differences in our methods of calculation. For example, Ref.~\cite{Huber:2000mg} assumes a thin-wall approximation and uses an ansatz for the profile that implicitly assumes small $\Delta\beta$, rather than solving for the full tunneling solution using path deformations as we do with {\tt CosmoTransitions}.  We also differ significantly from Ref.~\cite{Huber:2000mg} in our treatment of the stops \footnote{It is also worth noting that Ref.~\cite{Huber:2000mg} considers a different region of parameter space than we do}.

To illustrate the possible range of the one-step electroweak phase transition strength, as well as the parameters of $\Delta\beta$ and $L_w$, we perform a scan over values of $A_{\kappa}$  and for the three sets of points, Set I, II, III, corresponding to BM 1/2, 3/4, 5, respectively, keeping the rest of the parameter values fixed\footnote{Although the only benchmarks we consider with a strongly first order one-step EWPT are BM 1 and 2, all three sets of points we considered exhibit one-step electroweak phase transitions for some range of $A_{\kappa}$}.  As discussed in Sec.~\ref{sec:params}, the rest of the spectrum does not depend sensitively on $A_{\kappa}$ and so this amounts to varying the singlet-like CP-even/CP-odd masses.  The ranges of $A_{\kappa}$ considered in each case are $-A_{\kappa}\in [81 \hspace{.1cm} {\rm GeV}, 146 \hspace{.1cm}{\rm GeV}]$, $[87\hspace{.1cm} {\rm GeV}, 135\hspace{.1cm} {\rm GeV}]$, $[94\hspace{.1cm}{\rm GeV}, 146\hspace{.1cm}{\rm GeV}]$ for Set I, II, III, respectively.  The results of these scans are shown in Fig.~\ref{fig:results}.  The circle-, diamond-, and square-shaped points correspond to Sets I, II, III. 

From the standpoint of a one-step electroweak phase transition, the effects of varying $A_{\kappa}$ are clear: larger $\left|A_{\kappa}\right|$ results in a larger tree-level contribution to the barrier by the $\kappa A_{\kappa} s^3$ term in the effective potential.  Meanwhile, larger $\left|A_{\kappa}\right|$ results in a smaller $m_{h_s}$ for the rest of the parameters fixed.  This also explains the differences between the three curves: for a given singlet mass, Set I has the largest $|A_{\kappa}|$, while Set II has the smallest.  The effect of increasing $\left|A_{\kappa}\right|$ on $L_w$ is also rather straightforward to understand: a larger barrier results in a thinner wall (parametrically, $L_w \propto \Delta\phi / \sqrt{\Delta V}$, where $\Delta V$ represents an overall rescaling of both the barrier height and the difference in pressure between the two VEVs).  From Fig.~\ref{fig:results}, $\Delta\beta$ is larger for lighter singlet masses/stronger phase transitions.  This effect is less simple to understand, as it ultimately results from the complicated interplay of the various parameters in the potential.  However, we find a clear correlation between $\Delta\beta$ and the strength of the phase transition.  The effect on the wall velocity will be discussed in the next section.

The black points in Fig.~\ref{fig:results} do not have a runaway solution for the bubble wall equations of motion, and so may be viable for electroweak baryogenesis, as is the case for BM 1.  However, as advertised in Sec.~\ref{sec:runaway}, for very strong first order transitions, the bubble wall might not be slowed sufficiently by the plasma and can accelerate without bound.  Such is the case for the cyan-colored points in Fig.~\ref{fig:results}.  As an illustrative example of this runway case, we can consider BM 2, which corresponds to $A_{\kappa}=-129$ GeV in Set I.  The transition again proceeds in one step to the broken phase, as with BM 1.  The late-time bubble wall parameters are not well-defined in this case, since the wall may never enter a regime with constant velocity and friction.

The large value of $\left|A_{\kappa}\right|$ leads to a very strongly first-order transition, and inspection of the mean field and full effective potentials for this point in Fig.~\ref{fig:runaway} clearly shows the existence of a runaway solution. This suggests that BM 2 likely cannot result in successful electroweak baryogenesis \footnote{The existence of a runaway solution does not necessarily imply that the wall will in fact run away.  For example, Ref.~\cite{Konstandin:2010dm} showed that there can be sources of hydrodynamic obstruction which prevent the bubble from ever reaching the runaway regime.  We have checked against the criteria outlined in Ref.~\cite{Konstandin:2010dm}, and find that the hydrodynamic obstruction is negligible for BM 2.  This is because of the relatively large amount of supercooling for this point ($T_c=142.0$ GeV).  Thus, we expect that the bubble wall will indeed run away (although there may be other possible exceptions; see Ref.~\cite{Megevand:2013hwa}).}.  Nevertheless, fast moving bubble walls can be interesting from the standpoint of gravitational wave production. Table \ref{tab:gw} lists the values of $\alpha$ and $\zeta/H$, and the resultant amplitude $h^2 \Omega$ and peak frequency $\tilde{f}$ of gravitational waves produced by this strongly first-order transition, assuming that the bubble does in fact run away (see Sec.~\ref{sec:PT} for explanation of these quantities).  Unfortunately, the predicted spectrum is much too faint to be observed by Big Bang Observatory (BBO) \cite{Corbin:2005ny} or eLISA \cite{Binetruy:2012}. However,the spectrum of gravity waves we consider is only that coming from collisions of the bubbles and neglects other possibly important contributions from turbulence and other hydrodynamic effects \cite{  Kahniashvili:2008pf, Caprini:2009yp,  Hindmarsh:2013xza, Giblin:2014qia}.  Future work is required to assess whether our conclusions hold once these additional contributions are properly accounted for.  Note that the primary obstacle to achieving a detectable signature is the small size of $\alpha$. For points close to BM 2,5 with similar $T_n$, $d(S_3/T_n)/dT$ but larger $\alpha_N$ such that $K\approx 1$, the gravitational wave signatures could potentially be detectable by BBO, though likely not by eLISA.  Of course there may also be other regions of the NMSSM parameter space predicting much larger signals which would be interesting to explore in the future.  

\begin{figure*}[!t]\label{fig:runaway}
\subfigure{\includegraphics[width=0.4\textwidth]{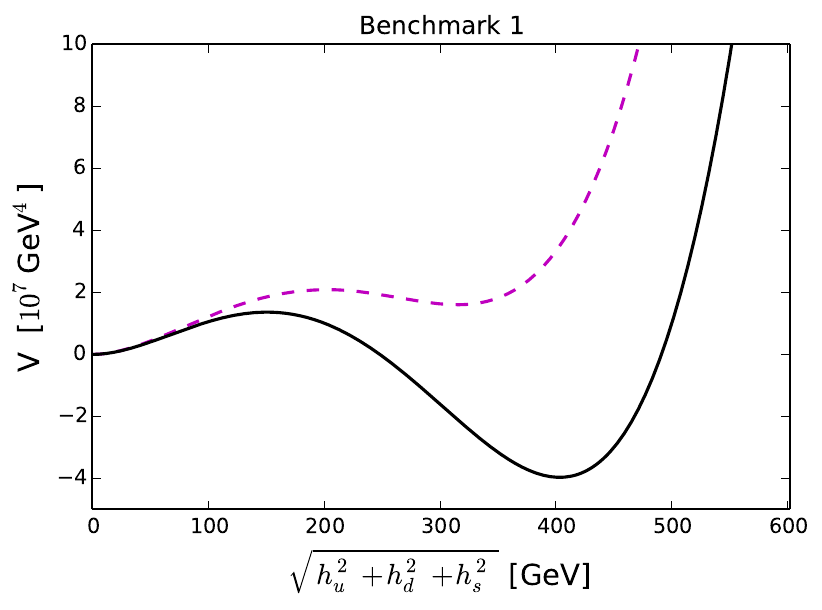}}
\subfigure{\includegraphics[width=0.4\textwidth]{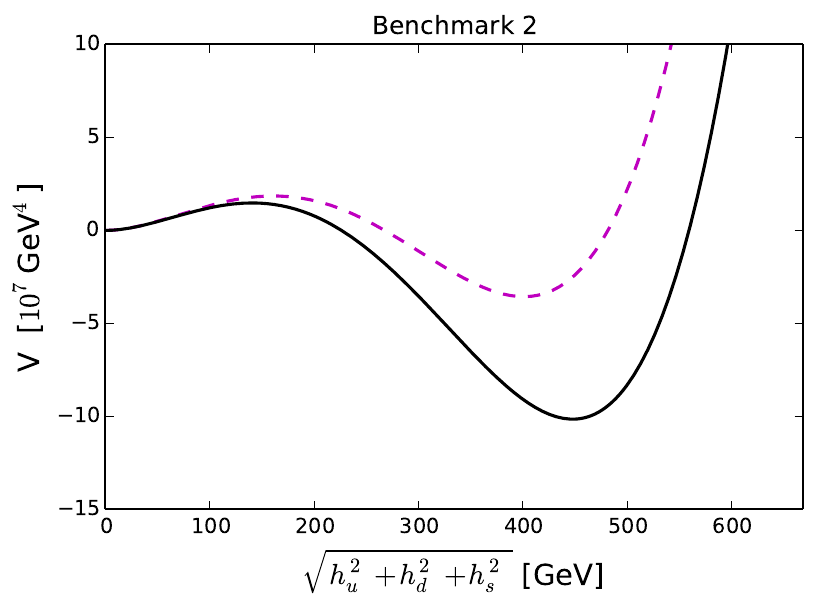}}

\subfigure{\includegraphics[width=0.4\textwidth]{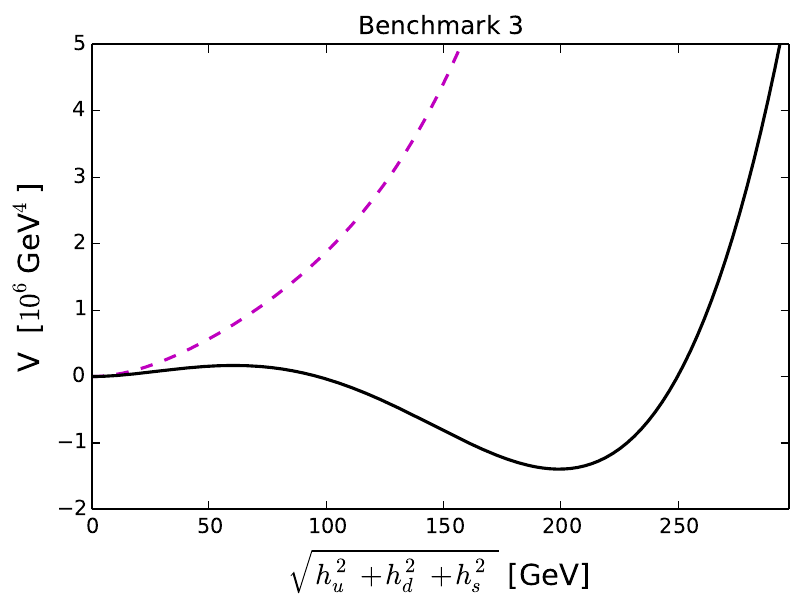}}
\subfigure{\includegraphics[width=0.4\textwidth]{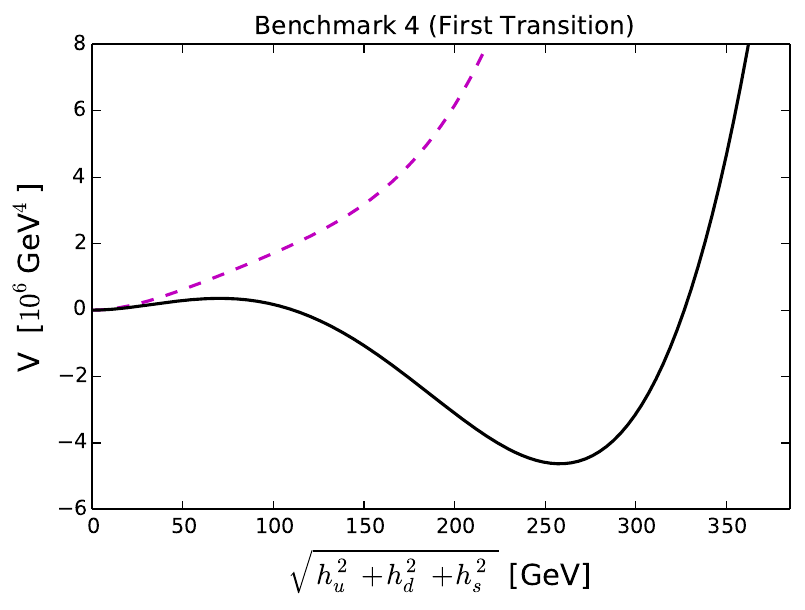}}

\subfigure{\includegraphics[width=0.4\textwidth]{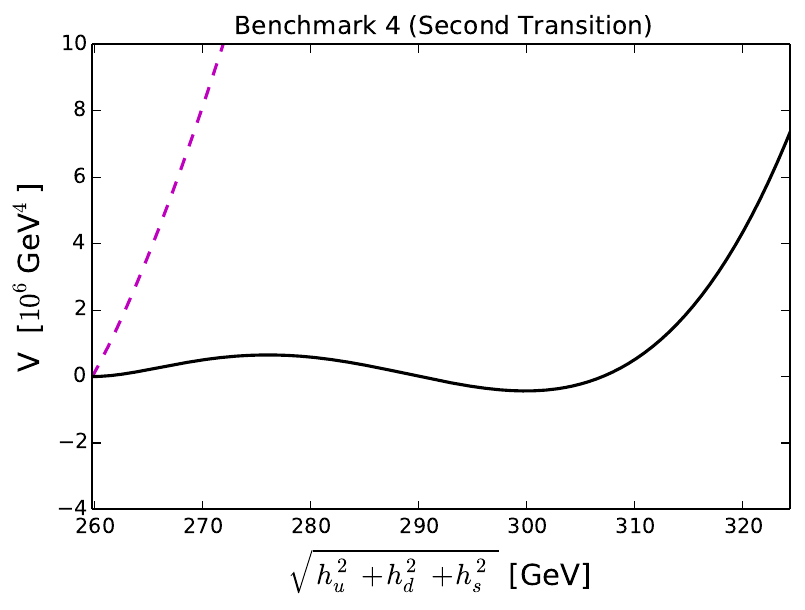}}
\subfigure{\includegraphics[width=0.4\textwidth]{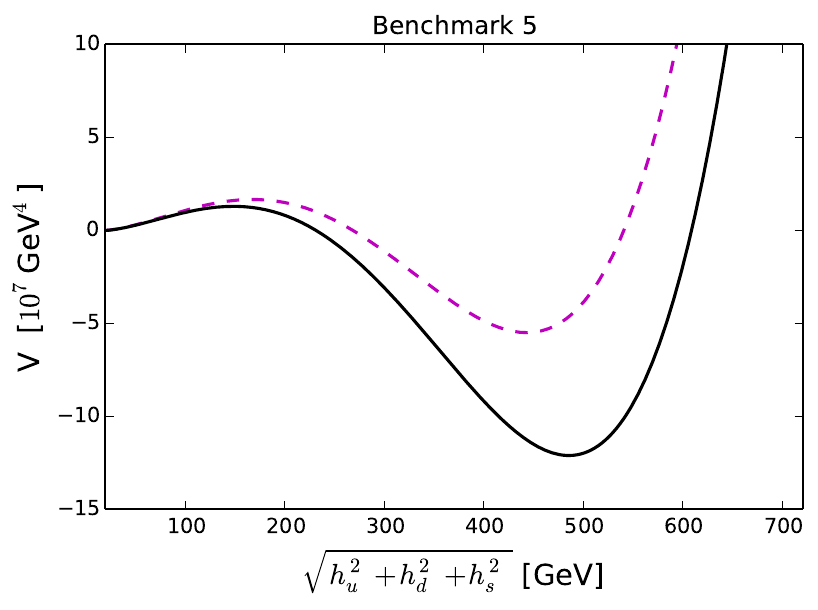}}

\caption{ \it \small Full (solid) and mean field (dashed) finite temperature 1-loop effective potential for all benchmarks at their respective nucleation temperatures.  Benchmarks 1, 3, and 4 have bubble walls which are guaranteed to remain sub-luminal, while the walls of BM 2 and 5 can have $v_w\rightarrow 1$.}

\label{fig:runaway}
\end{figure*}

In addition to the one-step EWPT case, there are several other patterns of symmetry breaking possible in the NMSSM.  Although not all are promising from the standpoint of electroweak baryogenesis, they may still yield interesting cosmological signatures.  These possibilities are illustrated by Benchmarks 3--5.

Benchmark 3 features a strongly first-order transition in only the singlet direction. This occurs at a nucleation temperature $T_n = 186.5\ \mathrm{GeV}$, while the critical temperature is $T_c=188.7$ GeV.  The amount of supercooling is significantly less than in the transitions involving the $SU(2)$ field directions.  The $SU(2)$ transition, which occurs at $T_n = 169.6\ \mathrm{GeV}$, is second-order and so is not shown in Table \ref{tab:outtab}. This is benchmark has the smallest value of $|A_\kappa|$, and correspondingly a thicker wall, with $L_s = 16.4/T_n$.  As can be seen from the third panel of Figure \ref{fig:runaway}, no runaway solutions exist.  We find this to be a generic feature of singlet-only transitions in the parameter space we have investigated: singlet-only transitions tend to have smaller total pressure difference than transitions in all three directions, since only one field is changing its value.

Another novel possibility in the NMSSM is that the phase transition can proceed in \emph{two steps}: a first transition results in a non-zero singlet VEV, while a second transition breaks electroweak symmetry.  This is exemplified by Benchmark 4. Here, the singlet transition occurs at $T_n=165.7$ GeV, while the $SU(2)$ transition is at a slightly lower temperature, $T_n=165.4$ GeV. For this particular benchmark, only the singlet phase transition is strongly first-order by our definition ($\Delta \phi/T_n>1$), while the transition in the $SU(2)$ directions is weakly first-order (has a value of $\Delta\phi/T_n<1$).  From the discussion in Sec.~\ref{sec:PT} concerning the baryon number preservation condition, future work is required to determine whether or not such a weak transition can lead to successful electroweak baryogenesis.  Regardless, this point serves as a proof of principle that the phase transition in the NMSSM can proceed in two steps. Regarding the bubble profile parameters, the wall width for the first transition is slightly thinner than that of BM 3, corresponding to the larger value of $\abs{A_\kappa}$. The second transition yields the thickest wall, simply because the phase transition is weak and the barrier between the two minima relatively low.  Note that $\Delta \beta$ is significantly smaller for this transition, with a value in the range typical for the MSSM \cite{Moreno:1998bq}.  This is expected, since the singlet is not participating in the transition.  Note that because of the relatively low pressure differences for each transition, neither permits runaway solutions, as shown in Fig.~\ref{fig:runaway}. 

Finally, Benchmark 5 produces a one-step strongly first-order transition at $T_n = 103\ \mathrm{GeV}$, and is unique among the benchmarks we considered in that $h_u$ has a nonzero high-temperature VEV. 
The symmetry is first broken by a second-order transition in the $h_u$ direction, and is then further broken in the remaining two directions when the first-order phase transition occurs at $T_n = 103$ GeV.  
Since electroweak symmetry is already partially broken in the phase outside the bubble, sphalerons will already be suppressed in this region, and so it is unlikely that successful electroweak baryogenesis will occur.  Also, from the sixth panel of Figure~\ref{fig:runaway}, we see that a runaway solution exists, making this transition, like that of BM 2, even less attractive for electroweak baryogenesis, but possibly interesting from the standpoint of detectable gravitational radiation (once again the hydrodynamic obstruction \cite{Konstandin:2010dm} is negligible due to the large supercooling, $T_c=141.2$ GeV). Table~\ref{tab:gw} shows that, like BM 2, BM 5 produces gravity waves with a peak frequency on the order of several mHz and an amplitude on the order of $10^{-17}$, which unfortunately lies below the range observable by BBO and eLISA \cite{Huber:2008,Binetruy:2012, Corbin:2005ny}.

\begin{table}
\begin{tabular}{c|c|c}
				&	BM 2					&	BM 5					\\
\hline
$\zeta/H$			&	670					&	1400					\\
$\alpha$			&	0.10				&	0.11				\\
$\tilde{f}$	 (mHz)	&	2.8					&	5.5				\\
$h^2 \tilde{\Omega}_\mathrm{GW}$		&\hspace{.1cm}	$2.1\times10^{-17}$  \hspace{.1cm}	& \hspace{.1cm}	$9.4\times10^{-18}$ \hspace{.1cm}	\\
\end{tabular}
\caption{\it \small Gravitational radiation spectra for the benchmarks with runaway solutions.  The predicted spectra fall well below the expected sensitivity of eLISA and BBO \cite{Binetruy:2012, Corbin:2005ny}.}
\label{tab:gw}
\end{table}

To summarize the results of this section, we have found that the very narrow region of parameter space investigated features a rich phase transition phenomenology.  Transitions can proceed in either one- or two-steps and can occur from either the origin in field space or from a phase with a non-zero VEV prior to tunneling.  Some transitions, namely those with a large amount of supercooling, produce bubble walls that can accelerate without bound; the gravitational radiation produced by these bubbles is however too faint to be observed by BBO or eLISA.  Most other transitions do not produce runaway bubble walls, and thus can give rise to successful electroweak baryogenesis.  The parameters governing the wall profile, such as $\Delta\beta$ and $L_w$, typically take on values more promising for electroweak baryogenesis than in the MSSM, making the NMSSM an even more attractive framework for simultaneously explaining the Higgs mass, dark matter, and baryogenesis.  

The only other important parameter not yet computed is the bubble wall velocity.  This is the task we turn to next.

\section{Estimating the Wall Velocity}\label{sec:vw}

We have seen in the previous section that many cases with a strongly first order electroweak phase transition predict bubble walls that approach a finite steady-state velocity.  However, the transport processes important for electroweak baryogenesis typically depend quite sensitively on the precise value of  $v_w$.  We would thus like to go beyond the analysis above and obtain a quantitative estimate of the $v_w$ for the scans shown in Fig.~\ref{fig:results}.   

Determining the wall velocity requires computing the drag force on the bubble wall, which is in general a difficult problem.  However, the situation is simplified in two limiting cases: the ultra- and non-relativistic (or ``slow-wall") limits. The former is the simplest, since the drag does not depend on the wall velocity or on the deviations from equilibrium of the various species in the plasma at lowest order in $1/\gamma$.  The friction saturates and if the driving force is greater than this value, the bubble wall runs away.  We considered this limit in our analysis of runaway solutions in Sec.~\ref{sec:runaway}.  In this Section we are interested in the opposite case: we will assume that the wall is propagating with velocity such that $\gamma\approx 1$.  In this regime we can estimate the friction force microphysically, as has been done for the SM \cite{Moore:1995si} and MSSM \cite{John:2000zq} some time ago.  This will allow us to determine the value of a phenomenological friction parameter $\Gamma$ that reproduces the approximate wall velocity, provided $v_w$ is not too large.

We emphasize that we \emph{start} with the assumption of a non-relativistic wall in our calculation.  Specifically, we assume a simple form for the friction coefficient in the bubble wall equations of motion that does not match on to the solution in the relativistic regime. Despite the shortcomings of this parametrization, it is nevertheless expected to provide a good estimate for the case of non-relativistic bubble walls \cite{Megevand:2013hwa}.  Consequently, we only compute the wall velocity for points that do not have runaway bubble wall solutions.  Furthermore, our results in this limit may be useful for more in-depth future studies of the wall velocity in the NMSSM, providing preliminary reference values to match on to the techniques of e.g. Refs.~\cite{Megevand:2013hwa, Huber:2013kj, Huber:2011aa}.  In particular, our computation of the various interaction rates of the particles in the plasma is quite general and can be used in future studies independent of the simplifying assumptions outlined above.  The reader should interpret our results with the above provisos in mind.

In the sub-sections below we discuss our computation of the stationary-state solution to the bubble wall equations of motion (Sec.~\ref{sec:SEOM}), as well as the microphysical friction coefficients (Secs.~\ref{sec:fric_sources},~\ref{sec:friction_coeff}; see also App.~\ref{ap:rates}).  Readers primarily interested in the results may proceed directly to Sec.~\ref{sec:vw_results}.

\subsection{Simplified equations of Motion}\label{sec:SEOM}
To find the bubble wall velocity in full generality, one must couple the field equations of motion to the radiation, and then also couple the radiation fluid equations to the field. The radiation will tend to slow down the bubble wall's expansion, whereas the forward motion of the wall will tend to heat the fluid and change its velocity. The effective potential is a function of both the field value and the temperature, so heating the fluid will change the free-energy (or pressure) difference between the two phases, thereby changing the bubble wall's acceleration. Additionally, the fluid velocity feeds back upon the wall by changing the magnitude of the drag. The interplay between these effects leads to a rich phenomenology: bubble walls can expand in steady detonations or deflagrations~\cite{Megevand:2009ut}, as runaway events~\cite{runaway}, or, if reheating is large enough, the two phases can reach pressure equilibrium and there can be an adiabatically varying period of phase coexistence~\cite{Wainwright:2009mq}.

Solving the full set of relativistic hydrodynamic equations is a difficult problem. However, we are primarily interested in rough estimates of the wall velocity, and are particularly interested in how the velocity compares to that in the MSSM light stop scenario. For this purpose, we ignore fluid velocity and temperature differentials across the bubble wall and their associated effects upon the bubble wall, and we treat the drag upon the wall as field-independent. Note that ignoring the shock front and temperature changes in the wall will tend to over-estimate the wall velocity for the deflagration case relevant for EWB \cite{Moore:1995si}.

With these simplifying assumptions, the field equations of motion are then
\begin{equation}
\partial^\mu \partial_\mu \phi_i + {\Gamma} u^\mu \partial_\mu \phi_i = -\partial_{\phi_i} V(\phi, T),
\end{equation}
where $u^\mu$ is the fluid four-velocity, and $\Gamma$ is the drag coefficient. The bubble will become thin-walled as it grows so we can approximate its motion as planar, and we assume that it reaches a steady velocity and a steady profile. The wall is static in its rest frame, and the equations of motion become
\begin{equation}
\label{eq:static_eom}
\frac{d^2 \phi}{dx^2} - \Gamma \gamma v_w \frac{d\phi}{dx} = \partial_{\phi_i} V(\phi, T).
\end{equation}
The velocity $v_w$ is the wall's velocity relative to the fluid. This is the quantity we wish to estimate. As usual, $\gamma = 1/\sqrt{1-v_w^2}$. If we center the bubble wall at $x=0$ in its rest frame, then the boundary conditions are $\phi(x=-\infty) = \phi_T$ (the field should be in the true vacuum deep inside of the bubble) and $\phi(x=\infty) = \phi_F$ (the field should be in the false vacuum far outside of the bubble). Also, $\frac{d}{dx}\phi(x=-\infty) = \frac{d}{dx}\phi(x=\infty) = 0$.

As mentioned in Sec.~IVA, we use Eq.~\ref{eq:static_eom} to solve for the late-time bubble profile parameters.  Assuming a common $\Gamma$ neglects the effects of the different drag forces experienced by the three fields. Improving on this simplified treatment is beyond the scope of our study, however, we believe that this approximation can give a reasonable estimate for the profile parameters.  The quantity $\Delta\beta$ will only be changed by the friction acting on the directions perpendicular to the path in field space, which are neglected in assuming a friction term proportional to $\phi'(x)$.  Differences in $\Gamma$ for the three fields will only act to change the field profiles in \emph{physical} space, since they act parallel to the field space trajectory.  This would correspond to a change in the relative size of $L_{u,d,s}$. Meanwhile, $\Gamma$ is calibrated to yield the same net drag force as that resulting from the microphysical friction parameters and thus provides a reasonable (rough) estimate of the size of the friction effects in the direction parallel to the path.  While the values of $\Delta\beta$ will change when the friction is modeled more accurately, we expect this correction to be relatively small to the extent that the perpendicular components of the friction are small.  Note that this assumption was also made in the calculation of the MSSM wall velocity found in Ref.~\cite{John:2000zq}.  While the changes in the wall widths may be somewhat larger, $L_{u,d,s}$ do not change significantly between their instanton and late-time values for most (thin-walled) points we consider, and so we expect this effect to be rather small as well.  Future work is required to precisely determine how large these effects might be.  Also, the precise form of the friction term used in Eq.~\ref{eq:static_eom} is not unique; including additional field-dependence, such as the $\phi^2$ term in Eq.~\ref{eq:EOM_1} below, will alter the precise shape of the bubble profiles.  A detailed treatment of this effect is beyond our scope here.  The reader should keep the above caveats in mind when interpreting our results.

Equation~\ref{eq:static_eom} is very similar to the equation governing the shape of the initial critical bubble, the only difference being that here the friction term is constant instead of inversely proportional to the bubble's radius. In one field dimension, we can solve for the boundary conditions using an overshoot/undershoot method similar to the one used for the critical bubble. When solving for the critical bubble, the friction term is fixed and the initial position $\phi(r=0)$ is varied to satisfy the boundary conditions ($\phi(r=\infty)=\phi_F$). When finding the velocity $v_w$, the initial position is fixed and the overall coefficient $\Gamma\gamma v_w$ can be varied to satisfy the boundary condition. If the coefficient is too low, the field will move too quickly ($|d\phi/dx|$ will be too large) and it will overshoot the false vacuum. Conversely, if the coefficient is too high, the field will not reach the false vacuum before $x=\infty$. Instead, $d\phi/dx$ will go to zero at finite $x$, and the field will eventually oscillate about the potential barrier maximum separating the two phases: an undershoot. By iterating between overshoots and undershoots one can converge upon the correct coefficient. We perform this iteration using {\tt CosmoTransitions}, but modified to vary the drag coefficient instead of $\phi(r=0)$. When the potential has multiple fields, the problem changes in exactly the same way for the critical bubble as for the drag coefficient calculation. We use the {\tt pathDeformation} module in {\tt CosmoTransitions} to find the correct path through field space.

With the above approach we can calculate the overall coefficient $\Gamma\gamma v_w$ for the various points of interest.  For example, $\Gamma \gamma v_w=27.24$ GeV, 9.11 GeV for benchmarks 1 and 3, respectively.  However, we would like to go beyond determining $\Gamma \gamma v_w$ and find the wall velocity itself. To do this, we need to calculate $\Gamma$.  We do so by matching onto a (simplified) microphysical calculation of the friction force in the non-relativistic limit, as described below.


\subsection{Sources of Friction}\label{sec:fric_sources}

Friction on the bubble wall arises from interactions of the wall with the various species in the plasma which dissipate the wall's energy.  Consequently, to determine $\Gamma$ in the non-relativistic regime, one must know not only the bubble properties, but also the out-of-equilibrium distributions of the various particles in the plasma that interact with the wall.  This is generally a very difficult problem, however it has been solved in several different approximation schemes for the SM and MSSM cases in the past (see e.g. Refs.~\cite{Moore:1995si, John:2000zq}).  In this Section, we will apply the techniques laid out in Refs.~\cite{Moore:1995si, John:2000zq} to calculate $\Gamma$ for our various benchmarks starting from the various interaction rates in the plasma with several simplifying assumptions. Our goal is not a precise calculation of $\Gamma$, but rather a quantitative estimate allowing us to compare the wall velocities across our benchmarks and to determine whether the bubble is likely to be sub-sonic or not.  

We begin with the equations of motion for a set of infrared (IR) scalar field condensates in a plasma. The scalar field Lagrangian and the conservation of energy result in a set of Klein-Gordon equations with damping terms for the fields $\phi_i$.  In the bubble wall frame, ignoring the curvature of the wall, we have \cite{Moore:1995si}
\begin{equation}
-\frac{d^2\phi_i}{dz^2}+\frac{\partial V(\phi_i,T)}{\partial \phi_i}+\sum_j\frac{d m_j(\phi_i)}{d \phi_i} \int \frac{d^3 p}{(2\pi)^3 2E}\delta f_j(p,z)=0
\end{equation}
where the index $j$ runs over all particles that couple to the scalar field $\phi_i$, and $\delta f_j$ connotes the deviation from the thermal equilibrium distribution of the particle species $j$ in the plasma, with 
\begin{equation}
f_j= \left(e^{\frac{E+\delta_j}{T}}\pm 1\right)^{-1}
\end{equation}
in the fluid frame.  In the above expression, the $\delta_j$ are generally spacetime-dependent perturbations from equilibrium generated by the interactions of the particle species $j$ with the bubble wall. Provided that the species $j$ satisfies the WKB condition $p_j\gg L_w^{-1}$ (we will deal with soft excitations below), the distributions satisfy Boltzmann equations (semi-classical versions of the Louville equations \footnote{For a derivation of the Boltzmann equations in the Schwinger-Keldysh formalism, see the recent Ref.~\cite{Konstandin:2014zta}}), given in the fluid frame by
\begin{equation}
\frac{\partial}{\partial t}f_j + \dot{x}\frac{\partial}{\partial x} f_j +\dot{p}_x\frac{\partial}{\partial p_x} f_j = -C[f_j]
\end{equation}
where $C[f]$ is a collision term which depends upon the various interaction rates of $j$ in the plasma:
\begin{equation}\label{eq:coll}
\begin{aligned}
C[f]=&\int \frac{d^3 kd^3p'd^3k'}{\left(2\pi\right)^92E_p 2E_{p'}2E_{k'}}\left|\mathcal{M}\right|^2(2\pi)^4\delta(p+k-p'-k')  \\
& \times f_pf_k(1\pm f_{p'})(1\pm f_{k'})-f_{p'}f_{k'}(1\pm f_p)(1\pm f_k).
\end{aligned}
\end{equation}  In the fluid approximation, the quantum mechanical field perturbations are assumed to take the form of a perfect fluid,
\begin{equation}
 \delta_j=-\left[\delta \mu_j+\frac{E}{T}\left(\delta T_i + \delta T_{bg}\right)+p_x\left(\delta v_i +\delta v_{bg}\right)\right],
\end{equation}
where the $bg$ subscript corresponds to the perturbations of the background fluid, assumed to comprise the degrees of freedom with small couplings to the Higgs fields.  Inserting this form of the perturbations back into the Boltzmann equation yields
\begin{equation}\label{eq:Boltz}
\begin{aligned}
&-f_0'\left(\frac{p_x}{E}\left[\partial_x \delta \mu_j + \frac{E}{T}\partial_x\left(\delta T_j+ \delta T_{bg}\right) + p_x \partial_x(\delta v_j + \delta v_{bg})\right]+\partial_t \delta \mu_j \right.\\&\left.+\frac{E}{T}\partial_t\left(\delta T_j+\delta T_{bg}\right)+p_x\partial_t\left(\delta v_j+\delta v_{bg}\right)\right)+T C\left[\delta \mu_j,\delta T_j,\delta v_j\right]=-f_0'\frac{\partial_t m^2\left(\phi\right)}{2E}
\end{aligned}
\end{equation}
These equations can be recast in the form \cite{Moore:1995si, John:2000zq}
\begin{equation}
\label{eq:vector}
\mathbf{A} \frac{d}{dx} \vec{\delta} + \mathbf{\Gamma} \vec{\delta} = \mathbf{F}
\end{equation}
where $\mathbf{A}$ is a matrix with entries $\sim c_{2,3,4} v_w$, $\mathbf{\Gamma}$ is a matrix involving the various interaction rates, $\mathbf{F}$ is the source term, and $\vec{\delta}$ is a vector comprising the various perturbations $\delta_i$.  After solving the above set of Boltzmann equations for the perturbations, the (space-time--dependent) solutions for the perturbations in the bubble wall can then be plugged into the Higgs equations of motion.  Using $\delta f_j \simeq f_0' \delta_j$, the equations of motion for the background Higgs fields become
\begin{equation}
-\phi_i''+\frac{\partial V}{\partial \phi_i}+\sum_j\frac{N_j}{2}\frac{dm^2_j(\phi)}{d\phi_i}\left(c_{1\pm}\delta \mu_j+c_{2\pm}\delta T_j+c_{2 \pm}T \delta v_j\right)=0,
\end{equation}
\begin{equation}
c_{n\pm}\equiv \int \frac{E^{n-2}}{T^{n+1}} f_0'(\pm)\frac{d^3p}{(2\pi)^3}.
\end{equation}

In general, one needs to solve the coupled set of Boltzmann equations represented by Eq.~\ref{eq:vector} to determine the perturbations $\delta_i$. However, the situation is simplified by noting that, when the wall velocity is small and the wall is not too thin, the terms involving $\delta_i'\sim \delta_i/L_w$ are multiplied by $c_{j} v_w$ and can thus be significantly smaller than the terms involving the $\delta_i$ (the latter are multiplied by the various rates which are typically of $\mathcal{O}(10^{-2}\, T)$ or larger).  Thus, in this regime, we can approximate the perturbations as roughly constant in the wall, $\delta'=0$.  Of course this approximation breaks down for faster moving, thinner walls, but we use it here to obtain a rough estimate to compare between our benchmarks and the MSSM.  Ref.~\cite{John:2000zq} compared the friction coefficients found using this approximation with the full numerical solution for the light stop MSSM scenario and found discrepancies up to a factor of 3 for the case of light stops.  When discussing the wall velocities for each of our benchmarks we address the implications of these possible differences and find that our overall conclusions remain unchanged.  

Under this simplifying assumption, the Boltzmann equations can be easily inverted for the perturbations $\delta_j$, resulting in $\vec{\delta}=\mathbf{\Gamma}^{-1} \mathbf{F}$.  Plugging in these solutions, the equations of motion become
\begin{equation}
\label{eq:EOM_1}
\phi_i''-\frac{\partial V(\phi,T)}{\partial \phi_i} = \eta_i v_w \gamma \frac{\phi_i^2}{T} \phi_i'
\end{equation}
where the $\phi_i'$ arises from the $\partial_t m^2(\phi)$ term on the RHS of Eq.~\ref{eq:Boltz}.  The $\eta_i$ are (constant) viscosity coefficients which characterize the friction from the plasma on the field direction $i$ of the expanding bubble wall.  These coefficients depend on the interactions of all species present in the plasma with one another and with the bubble wall and their calculation is quite difficult.  We dedicate Sec.~\ref{sec:friction_coeff} below to their computation.

The form of Eq.~\ref{eq:EOM_1} does not yet match that of the fluid equation (\ref{eq:static_eom}).  In particular, the damping term in Eq.~\ref{eq:EOM_1} carries additional space-time dependence by virtue of the $\phi^2$ term multiplying the derivative.  However, $\Gamma$ and $\eta$ can be easily related as follows \cite{Megevand:2003tg}: let us consider only one field direction for simplicity, with viscosity coefficient $\eta$ in Eq.~\ref{eq:EOM_1} and $\Gamma$ in Eq.~\ref{eq:static_eom} (the generalization to multiple field directions is given below).  Multiplying Eq.~\ref{eq:static_eom} by $\phi'$ and integrating over $x$ results in
\begin{equation}
\label{eq:etaGamma1}
\Gamma v_w \sigma = \Delta V
\end{equation}
where $\Delta V\equiv V(\phi_0,T_n)-V(\phi_n,T_n)$ is the pressure difference between the phases and 
\begin{equation}
\sigma\equiv \int (\phi')^2 dx
\end{equation}
is the surface tension of the bubble wall.  Performing the same integration on Eq.~\ref{eq:EOM_1} after multiplying by $\phi'$ yields
\begin{equation}
\label{eq:etaGamma2}
\Delta V =\frac{v_w \eta}{T}\int \phi^2 (\phi')^2 dx \simeq \frac{3 v_w \phi_n^2 \sigma \eta}{10T_n}
\end{equation}
where $\phi_n$ is the field value in the broken minimum at the nucleation temperature.  The last equality follows from assuming assuming a simple shape for the bubble wall profile 
\begin{equation}
\label{eq:profile}
\phi(x)=\frac{\phi_n}{2}\left(1+\tanh \frac{x}{L_w}\right).
\end{equation}
Finally, combining Eqs.~\ref{eq:etaGamma1} and~\ref{eq:etaGamma2} yields the desired relation between $\Gamma$ and $\eta$ for one field-dimension:
\begin{equation}
\label{eq:G_WKB}
\Gamma_{WKB}^{\rm one-dimension} \simeq \frac{3 \phi_n^2}{10 T_n} \eta.
\end{equation}
Thus, given the values of $\eta$ determined microphysically from the theory, along with the phase transition order parameter, one can determine the values of $\Gamma$ that enter into Eq.~\ref{eq:static_eom}. 

In our case, we require the generalization of Eq.~\ref{eq:G_WKB} to multiple fields.  In solving the simplified fluid equations analogous to Eq.~\ref{eq:static_eom} we assume a common $\Gamma$ for all field directions.  Of course the friction coefficients $\eta_{u,d,s}$ are different, since each field couples to different degrees of freedom.  However, we can determine an approximate value of $\Gamma$ that should produce the same wall velocity as that found by solving Eqs.~\ref{eq:EOM_1}.  This is done by carrying out the same procedure as for the single-field case, multiplying each equation by $\phi_i'$, integrating over $x$, and adding the three equations together, noting that 
\begin{equation}
\Delta V=\sum \int_{-\infty}^{\infty}dx \hspace{.1cm} \partial_{\phi_i}V(\phi_i,T) \phi_i '.
\end{equation}
Here $\Delta V$ is the pressure which must be balanced out by the friction force for a steady state bubble as before. Setting the wall velocities from both resulting equations equal to one another, we find
\begin{equation}\label{eq:Gamma_wkb_full}
\Gamma_{\rm WKB}=\frac{\sum \int_{-\infty}^{\infty}\eta_i\frac{\phi_i^2(x)}{T_n}\phi'^{2}(x) dx}{\sum \int_{-\infty}^{\infty}\phi_i'^{2}(x) dx}.
\end{equation}
If one assumes a simple hyperbolic tangent profile for the fields,
\begin{equation}
\label{eq:prof_simp}
\phi_i(x)\approx \frac{\Delta \phi_i}{2} \left(1+\tanh \frac{x}{L_i}\right)+\phi_{i,0},
\end{equation}
Eq.~\ref{eq:Gamma_wkb_full} can be approximated by
\begin{equation}
\label{eq:Gamma_wkb}
\Gamma_{\rm WKB}\approx \frac{\sum\eta_i \Delta \phi_i^2 \left(5\phi_{i,n}^2+5\phi_{i,0}^2-2\Delta \phi_i^2\right)L_i^{-1}}{10T_n\sum \Delta \phi_i^2 L_i^{-1}}
\end{equation} 
Here $\phi_{i,0}$, $\phi_{i,n}$ are the field values before and after tunneling, respectively, $L_i$ are the corresponding wall widths, and $\Delta \phi_i\equiv \phi_{i,n}-\phi_{i,0}$.  For the simplified cases of a singlet- or $SU(2)$-only transition from the origin, Eq.~\ref{eq:Gamma_wkb} simplifies further to
\begin{equation}
\Gamma_{\rm WKB} \approx \frac{3 }{10T_n} \left\{ 
\begin{array}{ll}
\vspace{.3cm} v_{s}^2 (T_n) \eta_s, & \hspace{.3cm} \mbox {singlet only} \\ 
   \vspace{.3cm} v^2(T_n) \left(\eta_u \sin^4\beta+\eta_d \cos^4\beta\right), &\hspace{.3cm}\mbox{$SU(2)$ only}
\end{array} \right.
\end{equation} 
where $\beta$ here is understood to be defined at the $T=T_n$ minimum, and we have assumed $L_u=L_d$ for the second case.  To obtain the above result we have assumed that $\tan\beta$ is constant along the tunneling path (i.e. ignored $\Delta \beta$), while in general it is space-time--dependent.  For our calculations of the wall velocity, we use the full late-time wall profile (via Eq.~\ref{eq:Gamma_wkb_full}) as computed by {\tt CosmoTransitions}.  Comparing these results with the simple hyperbolic tangent approximation via Eq.~\ref{eq:Gamma_wkb}, we find that the hyperbolic tangent approximation tends to underestimate the full result by a few percent but otherwise is in rather good agreement with the results obtained by using the full profile.  This approximation can thus be useful in future studies should the full profile not be computed.   

In addition to the damping from particles with $p\gg L_w^{-1}$ described above, infrared bosons will also contribute to the friction \cite{Moore:1996bn}.  At low momenta, the gauge fields can be treated by classical field theory and be shown to undergo over-damped evolution \cite{Moore:2000wx} (scalar fields are not over-damped and so their contribution is numerically much smaller; we ignore these contributions).  This gives rise to a viscosity coefficient that can be comparable to those obtained from the WKB contribution above\footnote{In fact, in the SM, this contribution can be shown to dominate.  However, the contribution from infrared gauge bosons scales as $\log[m_W (\phi) L_w]$ with $L_w$ decreasing for heavier Higgses.  Therefore, while the dominant contribution to the SM case, infrared bosons lead to a friction coefficient that is numerically smaller than that from the tops and non-IR bosons in our case.  Nevertheless, we include this contribution in what follows.}. The relevant $\Gamma$ terms have simple analytic expressions for the case of one field dimension \cite{Moore:2000wx},
\begin{equation} 
\label{eq:G_IR_1}
\Gamma^{\rm one-dimension}_{IR}\approx \frac{3T_n}{16\pi} g_2 \left(\frac{m_D(T_n)}{T_n}\right)^2 \log \left[m_W(\phi_n)L_w\right].
\end{equation}
where $m_D$ is the Debye mass of the $SU(2)$ gauge bosons.  Note that this expression is valid at leading log order, and hence there is an undetermined $\mathcal{O}(1)$ constant that will be added to the $log$ (all such non-logarithmic terms are dropped in this approximation).  In keeping with previous work, we neglect this constant term.  Eq.~\ref{eq:G_IR_1} generalizes easily to the case of three field directions as before:
\begin{equation}\label{eq:Gamma_IR_full}
\Gamma_{\rm IR}=\frac{\frac{3m_D^2 T_n}{32\pi} \int_{x_{\star}}^{\infty} \frac{\left[h_u(x)h_u'(x)+h_d(x)h_d'(x)\right]^2}{\left[h_u(x)^2+h_d(x)^2\right]^2 dx}}{\int_{-\infty}^{\infty} \left[h_u'^{2}(x)+h_d'^{2}(x)\right] dx}
\end{equation}
where $x_{\star}$ solves $L_w m_W(x_{\star})=1$ (this cuts off the logarithmic divergence in the numerator of Eq.~\ref{eq:Gamma_IR_full} and corresponds to the breakdown of the kinetic theory \cite{Moore:2000wx}).  Assuming a simplified hyperbolic tangent profile and neglecting $\Delta \beta$, Eq.~\ref{eq:Gamma_IR_full} simplifies to
\begin{equation}
\label{eq:G_IR_2}
\Gamma_{IR} \approx \frac{9m_D^2(T_n)T_n}{16\pi} \frac{1}{v^2(T_n)} \log \left[ \frac{g_2}{2}v(T_n) L_w \right]\left(\sin^4\beta+\cos^4 \beta\right)
\end{equation} 
Here again $\tan\beta$ is the ratio of $SU(2)$ Higgs VEVs at $T_n$.  Note that we did not have to solve any Boltzmann equations to determine $\Gamma_{IR}$ and so it is free of the uncertainty associated with our $\delta'=0$ approximation.  In what follows, we take the total friction coefficients $\Gamma$ to be sums of the $\Gamma_{WKB}$ and $\Gamma_{IR}$,
\begin{equation}
\label{eq:G_tot}
\Gamma_{\rm tot}= \Gamma_{\rm WKB}+\Gamma_{\rm IR}
\end{equation}
 which should be correct up to $\mathcal{O}(\alpha_w)$ \cite{Moore:1996bn, Megevand:2009ut}. 

Armed with expressions ~(\ref{eq:G_WKB}), (\ref{eq:G_IR_2}), and (\ref{eq:G_tot}), it remains to determine the friction coefficients $\eta_u$, $\eta_d$, $\eta_s$ for the fields $h_u$, $h_d$, and $s$, respectively.  This is the task we turn to next.

\subsection{The Friction Coefficients} \label{sec:friction_coeff}

We would like to evaluate the rates for various particle interactions in the plasma. In particular, the fields most relevant for the drag on the bubble wall are those with large couplings to the scalar fields.  In the Standard Model, the fields with the largest couplings to the Higgs are the top quarks, Higgs and $SU(2)$ gauge bosons.   In the NMSSM, the couplings of the neutralinos and charginos to the Higgses are also significant.  All fields with sizable couplings to $h_{u,d,s}$ can contribute substantially to the friction on the bubble wall.  To compute the friction precisely involves evaluating a large number of interaction rates which are, in general, space-time dependent, due to the changing VEVs.  This results in a complicated network of coupled Boltzmann equations represented schematically by Eq.~\ref{eq:vector}.  A full analysis of the drag force on the wall is beyond the scope of this paper; recall that we are already making several simplifications in the treatment outlined above. Thus, in what follows we will make several additional simplifying assumptions and approximations that will allow us to estimate the friction force.

For the $SU(2)$ Higgs fields $h_{u,d}$ we will retain the simplification employed in previous calculations \cite{Moore:1995si, John:2000zq} and neglect the friction exerted by Higgs bosons.  This was justified for the SM case because there there is only one Higgs degree of freedom (not counting the Goldstones) and for the EWPT to be strong required a Higgs lighter than the $W$ mass.  While neither is true in the NMSSM, including the Higgs contribution will only increase the friction.  Since we are mostly concerned with the bubble wall moving too quickly for efficient electroweak baryogenesis, ignoring the Higgs bosons in the friction for the $h_{u,d}$ fields will yield a conservative estimate of the drag, which is sufficient for our purposes.  Similarly, we will also ignore the neutralino and chargino contributions to the friction in the $SU(2)$ directions.  These approximations are not necessary but simplify our calculations significantly.

We will take the friction on $h_{u,d}$ as arising from the top quarks and $SU(2)$ gauge bosons, neglecting the $U(1)$ contribution.  All other species are treated as a common background (see Ref.~\cite{Moore:1995si} for details).  The $SU(2)$ gauge bosons are treated as one species $W$ with common chemical potential, as in Ref.~\cite{Moore:1995si}, with the masses, couplings, and rates averaged over the three physical fields.  Similarly, the left- and right-handed components of the tops, as well as anti-tops of both helicities, are considered as one species.  This corresponds simply to adding the relevant Boltzmann equations together, averaging, and multiplying by the number of degrees of freedom.  Thus, the rates simply add to one another.  Our set-up is precisely that of Refs.~\cite{Moore:1995si, John:2000zq} to which we refer the Reader for further clarification and discussion.

A novel feature of phase transitions in the NMSSM is that the singlet field $s$ is typically involved.  Thus, we need to compute the drag on this field direction as well.  Since $s$ is a singlet under all SM gauge groups, its only interactions are with the Higgs bosons, Higgsinos, and singlinos.  Once again, a proper calculation would involve many interactions with complicated matrix elements.  However, we can once again neglect the Higgs contribution and analyze only those coming from the Higgsino/singlino sector.  This will result in a conservative estimate of the friction (within our approximation scheme), since the drag provided by the Higgs bosons will add to that from the Higgsinos and singlinos.  The field-dependent masses for the Higgsinos/singlinos (in the gauge eigenstate basis) are given by
\begin{equation}
\begin{aligned}
m_{\widetilde{H}}^2(s)&\simeq \frac{1}{2}\left(\lambda ^2 s^2 \right)\\
m_{\widetilde{S}}^2(s)&\simeq \frac{1}{2}\left(\kappa^2 s^2\right)
\end{aligned}
\end{equation}
The contribution of these fields to the friction on $s$ is proportional to $d m_i^2(s)/ds$ and so the contribution of the singlet and singlino to the friction will be suppressed by $\kappa^2/\lambda^2$, which is small for our benchmarks.  Therefore we can drop these contributions and consider only friction arising from the Higgsinos. 

We need to calculate the various interaction rates for the tops, gauge bosons, and Higgsinos which enter the matrix $\mathbf{\Gamma}$ discussed in the previous subsection.  The precise definition of $\mathbf{\Gamma}$ in terms of the various interaction rates is given in Ref.~\cite{John:2000zq}.  The quantities we need for each species $i$ are given by integrals of the collision term in Eq.~\ref{eq:coll}:
\begin{equation}
\begin{aligned}
\int \frac{d^3 p}{(2\pi)^3 T^2} C[f] = &\delta \mu_i \Gamma_{\mu_1,i}+\delta T_i \Gamma_{T_1,i}\\
\int \frac{d^3 p}{(2\pi)^3 T^3}E_i C[f] = &\delta \mu_i \Gamma_{\mu_2,i}+\delta T_i \Gamma_{T_2,i}\\
\int \frac{d^3 p}{(2\pi)^3 T^4} p_{x,i} C[f] = &\delta v_i \Gamma_{v_1,i}
\end{aligned}
\end{equation}
Thus, we require the various matrix elements for the processes contributing to $C[f]$.  

The relevant processes include contributions from diagrams with the $t$-channel exchange of a potentially soft gauge boson. At finite temperature, the contributions from soft bosonic degrees of freedom must be treated with some care, since, naively, they result in infrared divergences in the massless limit.  In reality, the divergence is cut off by the thermal self-energy in the propagator, which results instead in a finite, large logarithm $\sim \log1/\alpha$.  Consequently, the first analysis \cite{Moore:1996bn} of the Standard Model case utilized a ``leading logarithmic expansion" of the interaction rates, in which only $t$-channel processes are kept and the external particles were treated as massless.  Additionally, Ref.~\cite{Moore:1996bn} replaced the full thermal self-energies of all massless exchanged particles with the corresponding thermal/Debye masses.  The phase space integrals of Eq.~\ref{eq:coll} were then performed analytically.  In Ref.~\cite{John:2000zq}, a similar approximation was made, with the additional improvement that the integrals were performed numerically.  Subsequent analyses in different contexts \cite{Arnold:2000dr, Arnold:2003zc,Moore:2001fga} have improved upon these methods, and we draw on these analyses in our treatment of the rates.

Specifically, we consider all leading log order contributions to the various interaction rates, systematically dropping terms of order $m/T$, as in Refs.~\cite{Moore:1996bn, John:2000zq}.  However, in the case of the Higgsinos, there are also important $s$-channel contributions which are e.g. enhanced by $N_f$, the number of fermions or by the top Yukawa coupling and $N_c$.  These contributions, although formally higher in the logarithmic expansion, will contribute comparably to the $t$-channel pieces (see Appendix~\ref{ap:rates}).  We therefore include these contributions as well.  Following the results of Refs.~\cite{Arnold:2000dr, Moore:2001fga, Arnold:2003zc}, the matrix elements entering the collision integral are computed in terms of Mandelstam variables at zero temperature, and subsequently `translated' into the appropriate form for the finite temperature results, i.e. including the thermal self energies in the propagators, using the dictionary provided in Refs.~\cite{Arnold:2000dr, Moore:2001fga, Arnold:2003zc}.  The relevant vacuum matrix elements are given in Appendix~\ref{ap:rates}.  For the evaluation of the phase space integrals in Eq.~\ref{eq:coll}, we use the hard thermal loop approximation for the full thermal self-energies for all exchanged particles.  These are again detailed e.g. in Ref.~\cite{Arnold:2003zc}.  We evaluate the phase space integrals numerically using the \texttt{Cuba} Monte Carlo integration package \cite{Hahn:2004fe}.  The evaluation of these integrals is straightforward and is detailed in Refs.~\cite{Arnold:2000dr, Moore:2001fga, Arnold:2003zc}, along with the relevant expressions for the thermal self-energies.  Further details can be found in Appendix~\ref{ap:rates}.

For points of interest we typically find $\eta_u\sim 6$, $\eta_d\sim 0.3$, $\eta_s\sim 3$.  The $u$ and $d$ values are similar to those found in Ref.~\cite{John:2000zq} for a plasma with Standard Model particle content.  The difference is due to the full inclusion of the hard thermal loop self-energies in the $t$-channel propagators, which Ref.~\cite{John:2000zq} neglects.  The large values of $\eta_s$ are due to the smallness of the Higgsino interaction rates, which are suppressed due to the lack of colored interactions and, in the absence of light sfermions, the lack of leading log processes enhanced by $N_f$, the number of $SU(2)$ doublet Standard Model fermions.

\subsection{Approximate results for $v_w$}\label{sec:vw_results}

\begin{figure*}[!t]
\mbox{\includegraphics[width=0.45\textwidth,clip]{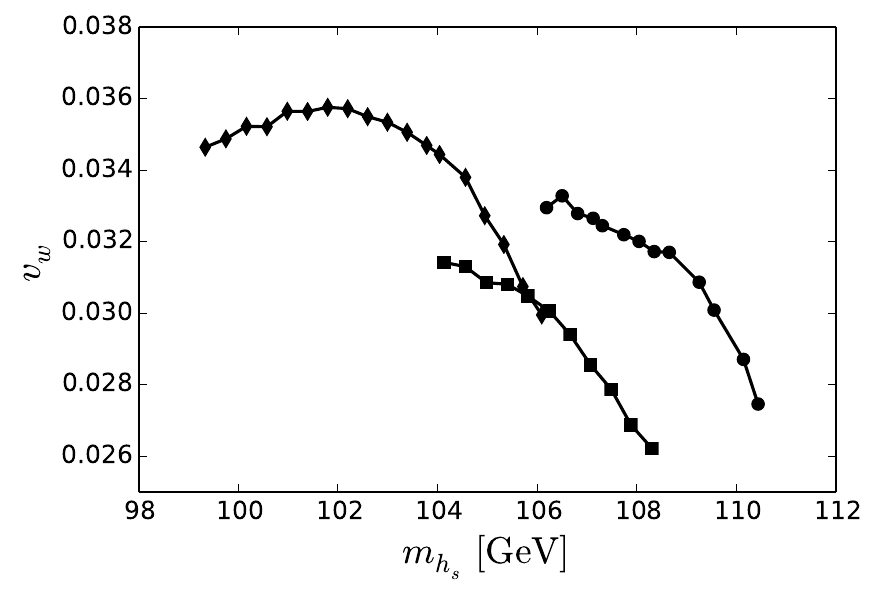}}\caption{\label{fig:vwall} \it \small Estimated wall velocities for the non-runaway points in Sets I, II, III (circles, diamonds, squares).  Care should be taken in interpreting these results in light of the approximations and simplifications we have made (detailed in the text), namely utilizing a simple form for the friction coefficient in the bubble wall equations of motion, and neglecting the spatial variation of the perturbations in the bubble wall.  A likely increase in $v_w$ of factor of a few relative to the values shown should be kept in mind for all cases.  However, even with this modification, all points considered appear to be sub-sonic, as required for successful transport-driven baryogenesis.}
\end{figure*}

With the interaction rates evaluated, we can now compute the wall velocities for all points in the scans shown in Fig.~\ref{fig:results}, given our assumptions.  We only consider points without runaway solutions.  The results are shown in Fig.~\ref{fig:vwall}.  The curves of Fig.~\ref{fig:vwall} demonstrate the expected parametric behavior: stronger transitions occur for smaller values of $m_{h_s}$, which therefore tend to have larger pressure and hence larger wall velocities.  The difference between the three sets of points is due to variations in the quantity $\Gamma \gamma v_w$, which is related to the pressure difference driving the wall expansion (see Eq.~\ref{eq:etaGamma1}).  For a given $m_{h_s}$, Set III has the smallest $\Gamma \gamma v_w$ and hence the smallest wall velocity (the drag coefficients $\eta_{u,d,s}$ are similar for the three sets considered). 

We emphasize that the velocities we have computed are estimates that likely substantially \emph{underestimate} the full result, given by including the $\delta'$ term in Eq.~\ref{eq:vector}.  This was pointed out in Ref.~\cite{John:2000zq} for the case of the MSSM, in which case the $\delta'=0$ approximation underestimated the wall velocity by a factor of 3 for the case of light stops.  Indeed, the results shown in Fig.~\ref{fig:vwall} do not match on to the velocities in the relativistic regime.  It is thus likely that some of the points with smaller $m_{h_s}$ feature significantly higher velocities, since they should match on to the points with runaway bubble walls.  Dropping the derivative term in Eq.~\ref{eq:vector} is an even more crude approximation for the Higgsinos than for the (s)tops, since the typical Higgsino interaction rates in the plasma are small (there are no interactions involving color charges or enhanced by $N_f$ at leading log order). However, we have checked that even completely neglecting the friction from the Higgsinos the computed wall velocity is subsonic for most points, which is encouraging. Still, a full treatment of the Boltzmann and wall equations is necessary to accurately determine $v_w$, and we hope to address this in future work.  

Keeping the above remarks in mind, our rough estimates preliminarily suggest that the effect of friction in the NMSSM might be large enough to cause the bubble walls to move sub-sonically (in the non-runaway cases).  More work is needed to verify this conclusion, however, if it were true, it would indicate that successful transport-driven electroweak baryogenesis could occur in the NMSSM parameter space considered.  Even an order of magnitude increase in the wall velocities shown in Fig.~\ref{fig:vwall} would not change this conclusion.

\section{Summary and Conclusions} \label{sec:conc}

In this paper we analyzed the nature and properties of the electroweak phase transition in the next-to-minimal supersymmetric Standard Model in light of the Higgs discovery at the LHC. We honed in on a region of parameter space featuring significant tree-level contributions to the Higgs mass, a viable dark matter candidate in the lightest supersymmetric particle (the lightest neutralino), stops in the TeV range, and the rest of the particle spectrum compatible with LHC searches. We employed an effective field theory approach to carefully compute the finite temperature effective potential, which was then fed to the {\tt CosmoTransitions} software package to study the details of symmetry breaking and compute the values of several parameters crucial for the calculation of the baryon asymmetry of the universe.

We showed that the phase transition structure, for phenomenologically viable parameter space points, is very rich.  There can be one-step electroweak phase transitions from the origin in field space or from some non-zero VEV prior to tunneling, one-step singlet-only transitions (which might give rise to observable gravitational radiation in some other regions of parameter space), as well as two-step phase transitions.  Although some of these possibilities seem not to lead to successful electroweak baryogenesis, such a major event in the history of the universe is certainly interesting cosmologically.  We believe it would be worthwhile to study the potential observational probes of such scenarios in the hopes of disentangling the NMSSM (or other theories with additional gauge singlets) from more minimal models.  

In addition to computing the patterns of symmetry breaking, a major aim of our study was investigating the microphysical parameters that enter into any realistic calculation of the baryon asymmetry via electroweak baryogenesis.  To this end, we studied the bubble wall profile, and in particular the quantities $L_w$, $L_s$, and $\Delta\beta$.  These parameters can vary to up to an order of magnitude even across the narrow slices of parameter space around our benchmarks.  Notably, we found that these parameters tend to take on values more promising for electroweak baryogenesis than in the MSSM, further suggesting the viability of NMSSM electroweak baryogenesis.

A crucial part of our study comprised our analysis of the bubble wall velocity for realistic parameter space points.  We found that ultra-relativistic solutions to the bubble wall equations of motion exist typically when the phase transition is very strong (when there is substantial supercooling).  For weaker transitions the bubble wall velocity tends to be sub-luminal, and thus potentially compatible with successful electroweak baryogenesis.  We provided an approximate estimate of the bubble wall velocity in the non-runaway case, hinging upon a microscopic treatment of the various sources of friction acting on the expanding bubble wall.  Although this estimate is rather rough in many respects, our results suggest typical values for the wall velocity are in the $\mathcal{O}(0.01-0.1)$ range, comparable to that of the MSSM light stop scenario.  Future work is required to improve this estimate. However, we stress that our computation of the various interaction rates of the particles in the plasma was quite thorough, and can be used in future studies beyond the simple framework we have employed here.

We believe this study can serve as a jumping off point for more detailed investigations of NMSSM electroweak baryogenesis, especially as the LHC, dark matter searches, and intensity frontier experiments continue to clarify what physics might exist beyond the Standard Model. Overall, our results suggest that the NMSSM might not only explain the observed Higgs mass, the nature of dark matter, and alleviate the hierarchy problem, but also explain the origin of the baryon asymmetry of the universe via electroweak baryogenesis. 

\begin{acknowledgments}
\noindent  We are very grateful to Stephan Huber and David Morrissey for many useful discussions and clarifications.  JK also thanks Travis A. W. Martin for input on the LHC phenomenology of the benchmarks.  JK and SP also thank the Kavli Institute for Theoretical Physics for their hospitality while the final portion of this work was being completed.  JK is supported by the Natural Sciences and Engineering Research Council of Canada (NSERC).  SP and LSH are partly supported by the US Department of Energy under Contract DE-FG02-04ER41268.  This research was also supported in part by the National Science Foundation under Grant No. NSF PHY11-25915.
\end{acknowledgments}

\section*{Note Added}

While the final stages of this work were being completed, Ref.~\cite{Huang:2014ifa} appeared, which investigates phase transitions in a similar region of the NMSSM.  They reach conclusions similar to ours in terms of the possible patterns of symmetry breaking.  Their study differs substantially from our investigation in several respects.  In particular, they include the stops as dynamic degrees of freedom in the one-loop zero-temperature effective potential, while we integrate them out and work in an effective field theory below the stop mass scale.  Also, they analyze the effective potential analytically to determine the conditions necessary for two of the vacua to be degenerate at non-zero temperature, as well as scan over the parameter space to find regions satisfying this condition, while our study is much more focused on the precise details of the phase transition with respect to the tunneling solution, bubble wall profile, and dynamics of the bubble wall expansion, which they do not attempt to address.  In this sense, their study is largely complementary to ours, but quite disjoint.

\appendix

\section{Renormalization Group Equations}\label{ap:RGEs}

In this appendix we list the one-loop renormalization group equations for the couplings that enter the effective 2HD+S potential.  The RGEs are valid below the stop mass scale and above the top mass scale.  We include the effects of the top quark, gauge bosons, Higgs/singlet bosons, Higgsinos, and singlinos.  The gaugino contributions are not included, since we take $M_2$ to be rather large and the bino contribution is numerically small.

The beta functions are defined as
\begin{equation} \label{eq:beta}
\beta_{p_i} = \frac{\partial}{\partial \log \Lambda^2} p_i
\end{equation}
where $p_i$ is the parameter of interest.  The RGEs for the various quartic couplings are:
\begin{equation}
\begin{aligned}
16\pi^2 \beta_{\lambda_1}=&6 \lambda_1^2+2\lambda_3^2+2\lambda_3\lambda_4+\lambda_4^2+\lambda_5^2-\frac{1}{2}\lambda_1(3g_1^2+9g_2^2) +\frac{3}{8} g_1^4+\frac{9}{8}g_2^4+\frac{3}{4}g_1^2g_2^2\\
&-2\widetilde{\lambda}^4+2\widetilde{\lambda}^2\lambda_1\\
16\pi^2\beta_{\lambda_2}=&6\lambda_2^2+2\lambda_3^2+2\lambda_3\lambda_4+\lambda_4^2+\lambda_6^2-\frac{1}{2}\lambda_2(3g_1^2+9g_2^2)+\frac{3}{8}g_1^4+\frac{9}{8}g_2^4+\frac{3}{4}g_1^2g_2^2 \\
&+6y_t^2\lambda_2-6y_t^4-2\widetilde{\lambda}^4+2\widetilde{\lambda}^2\lambda_2 \\
16\pi^2\beta_{\lambda_3}=&(\lambda_1+\lambda_2)(3\lambda_3+\lambda_4)+2\lambda_3^2+\lambda_4^2+\lambda_5\lambda_6-\frac{1}{2}\lambda_3(3g_1^2+9g_2^2)+\frac{3}{8} g_1^4+\frac{9}{8}g_2^4-\\
&\frac{3}{4}g_1^2g_2^2+3y_t^2\lambda_3-2\widetilde{\lambda}^4+4\widetilde{\lambda}^2\lambda_4+4\widetilde{\lambda}^2\lambda_3\\
16\pi^2\beta_{\lambda_4}=&\lambda_4(\lambda_1+\lambda_2+4\lambda_3+2\lambda_4)+2\lambda_7^2-\frac{1}{2}\lambda_4(3g_1^2+9g_2^2)+\frac{3}{2}g_1^2g_2^2\\
&+3y_t^2\lambda_4+2\widetilde{\lambda}^4-2\widetilde{\lambda}^2\lambda_4\\
16\pi^2\beta_{\lambda_5}=&\lambda_5(3\lambda_1+2\lambda_5+4\lambda_8)+\lambda_6(2\lambda_3+\lambda_4)+4\lambda_7^2-\frac{1}{4}\lambda_5(3g_1^2+9g_2^2)\\
&-8\widetilde{\kappa}^2\widetilde{\lambda}^2-2\widetilde{\lambda}^4+2\widetilde{\kappa}^2\lambda_5+3\widetilde{\lambda}^2\lambda_6\\
16\pi^2\beta_{\lambda_6}=&\lambda_5(2\lambda_3+\lambda_4)+\lambda_6(3\lambda_2+2\lambda_6+4\lambda_8)+4\lambda_7^2-\frac{1}{4}\lambda_6(3g_1^2+9g_2^2)\\
&+3y_t^2\lambda_6-8\widetilde{\kappa}^2\widetilde{\lambda}^2-2\widetilde{\lambda}^4+2\widetilde{\kappa}^2\lambda_6+3\widetilde{\lambda}^2\lambda_6\\
16\pi^2\beta_{\lambda_7}=&\lambda_7(\lambda_3+2\lambda_4+2\lambda_5+2\lambda_6+2\lambda_8)-\frac{1}{4}\lambda_7(3g_1^2+9g_2^2)\\
&+\frac{3}{2}y_t^2\lambda_7+4\widetilde{\kappa}\widetilde{\lambda}^3+2\widetilde{\kappa}^2\lambda_7+3\widetilde{\lambda}^2\lambda_7 \\
16\pi^2\beta_{\lambda_8}=&\lambda_5^2+\lambda_6^2+2\lambda_7^2+10\lambda_8^2+(4\widetilde{\kappa}^2+4\widetilde{\lambda}^2)\lambda_8-8\widetilde{\kappa}^2-2\widetilde{\lambda}^4.
\end{aligned}
\end{equation}
The RGEs for the dimensionful parameters $m_4$ and $m_5$ are given by
\begin{equation}
\begin{aligned}
16\pi^2\beta_{m_4}=&(\lambda_3+2\lambda_4+\lambda_5+\lambda_6+2\widetilde{\lambda}^2+\widetilde{\kappa}^2 - \frac{9}{4}g_2^2-\frac{3}{4}g_1^2+\frac{3}{2}y_t^2)m_4+2\lambda_7 m_5\\
16\pi^2\beta_{m_5}=&(6\lambda_8+3\widetilde{\lambda}^2+3\widetilde{\kappa}^2)m_5+6\lambda_7 m_4.
\end{aligned}
\end{equation}

In the above equations, $\widetilde{\lambda}$ and $\widetilde{\kappa}$ are the parameters appearing in the neutrlalino/chargino mass matrices.  We do not include their running, since they enter only in the one-loop contributions to the effective potential, and thus their running is formally a higher-order effect.  The same reasoning holds for the Yukawa and gauge couplings.  It should also be noted that the NMSSM parameters we match onto at the stop scale are technically $\overline{\rm DR}$ running parameters (appropriate for a supersymmetric theory), while the parameters we use in the effective theory are defined in the $\overline{\rm MS}$ scheme.  Converting between the two schemes results in a small threshold correction to several of the quartic couplings.  However, this only affects the quartics at the $\mathcal{O}(1\%)$ level, and so we neglect these corrections in our calculations.  

In practice, to match onto the spectrum calculated by \texttt{NMSSMTools}, when computing the corrections to the 2HD+S potential parameters we consider only the top, gauge boson, Higgsino, and singlino contributions, since \texttt{NMSSMTools} does not include Higgs boson corrections to the lightest CP-even state \cite{NMSSMTools}.  We have modified \texttt{NMSSMTools} accordingly so that our spectra exhibit good agreement.

\section{Vacuum Matrix Elements for Annihilation and Scattering Rates}\label{ap:rates}
In this Appendix we list the vacuum squared matrix elements, $\left| \mathcal{M}\right|^2_{ij\rightarrow kl}$, relevant for computing the friction coefficients for the fields $h_{u,d,s}$.  We sum over colors, spins, polarizations, and particle-antiparticle of all four states, then divide through by the number of degrees of freedom for the relevant incoming particle.  This convention matches that appearing in previous calculations of the matrix elements for the SM \cite{Moore:1995si} and MSSM \cite{John:2000zq} friction cases, although here we also include the largest contributions beyond the leading-log approximations of Ref.~\cite{Moore:1995si}.  Processes which do not contribute at LLO are marked with an asterisk, corresponding to the $N_f$--enhanced $s$-channel Higgsino pair annihilation contribution and Higgsino-singlino co-annihilation, which enters with large coupling $\lambda^2y_t^2$.     

\subsection{Friction on $s$}
For the CP-even singlet field, the relevant interactions are those that involve Higgsino (co-)annihilation and scattering.  The dominant contributions that we take into account (discussed in Sec.~\ref{sec:friction_coeff}) are

\begin{center}\uline{Higgsino Annihilation:} \end{center}
\vspace{-.5cm}
\begin{align}
\widetilde{H}\widetilde{H}\rightarrow W W: \hspace{1cm} &\frac{9}{8}g_2^4\left(\frac{u}{t}+\frac{t}{u}\right) \\
(*) \hspace{.2cm}  \widetilde{H}\widetilde{H}\rightarrow f \bar{f}: \hspace{1cm} &18 g_2^4\frac{u^2+t^2}{s^2}\\
(*) \hspace{.2cm}  \widetilde{H}\widetilde{S}\rightarrow t \bar{t}: \hspace{1cm} &3\lambda^2 y_t^2
\end{align}

\begin{center}\uline{Higgsino Scattering:} \end{center}
\vspace{-.5cm} 
\begin{align}
\widetilde{H}f \rightarrow \widetilde{H} f: \hspace{1cm} &18g_2^4\frac{s^2+u^2}{t^2} \\
\widetilde{H} W \rightarrow \widetilde{H} W: \hspace{1cm} &6 g_2^4\frac{s^2+u^2}{t^2}
\end{align}

Although we do not utilize them in this work, we also list the relevant Higgs interaction matrix elements for use in future studies:

\begin{center}\uline{Higgs Annihilation:} \end{center}
\vspace{-.5cm}
\begin{align}
H g\rightarrow t \bar{t}: \hspace{1cm} &8g_3^2 y_t^2 \frac{u}{t} \\
 H t\rightarrow t g: \hspace{1cm} &8g_3^2 y_t^2 \frac{u}{t} \\
H H \rightarrow t\bar{t}:  \hspace{1cm} &9  y_t^4 \frac{u}{t}
\end{align}

\begin{center}\uline{Higgs Scattering:} \end{center}
\vspace{-.5cm}
\begin{align}
 H f\rightarrow Hf: \hspace{1cm} &18g_2^4  \frac{su}{t^2} \\
 H t\rightarrow t H: \hspace{1cm} &9 y_t^4 \frac{s}{t} \\
H W \rightarrow HW:  \hspace{1cm} &\frac{15}{4} g_2^4 +3 g_2^4 \frac{s^2-st}{t^2}.
\end{align}

The above squared matrix elements are those relevant in the unbroken $SU(2)$ phase.  Once the exchanged particles obtain a non-zero mass term in the Lagrangian, new channels are opened which involve a mass insertion on the internal propagator, as well as several interference terms which vanish in the massless limit.  Additional contributions to the scattering rates tend to lower the viscosity, as they damp the perturbations of the equilibrium distribution functions. However, these contributions are typically suppressed by powers of $m/T$ and should only change the scattering rates significantly relatively far into the bubble wall.  Furthermore, we are only interested in a rather rough estimate of the wall velocity in each case.  We thus only consider the rates relevant for massless particles, noting that a full calculation should go beyond this approximation.  We do not expect these additional processes to significantly alter our results.  Note that this treatment is in keeping with that of Refs.~\cite{Moore:1995si, John:2000zq}. 

\subsection{Friction on $h_{u,d}$}

The relevant processes we consider in this case are annihilation and scattering of top quarks and $SU(2)$ gauge bosons.  These are the interactions treated in Ref.~\cite{Moore:1995si} for the SM case.  As discussed in Sec.~\ref{sec:friction_coeff}, we neglect the contribution of the Higgses, Higgsinos, and gauginos on the $SU(2)$ field-directions, noting that they will decrease the wall velocity when included.  Our treatment is thus conservative from the standpoint of electroweak baryogenesis, which requires sub-sonic wall velocities.

We include only those processes contributing large logs to the interaction rates. However, different from Refs.~\cite{Moore:1995si, John:2000zq}, we employ a full leading-order treatment in evaluating these contributions, instead of making the analytic approximations found in Ref.~\cite{Moore:1995si} or using only the numerical phase space integration of Ref.~\cite{John:2000zq}.  The resulting squared matrix elements (again summing over all degrees of freedom and then dividing out by the degrees of freedom of the incoming particle), are

\begin{center}\uline{Top Quark Annihilation:} \end{center}
\vspace{-.5cm}
\begin{align}
t \bar{t} \rightarrow gg: \hspace{1cm} &\frac{32}{9}g_3^4\left(\frac{u}{t}+\frac{t}{u}\right)
\end{align}

\begin{center}\uline{Top Quark Scattering:} \end{center}
\vspace{-.5cm} 
\begin{align}
t q \rightarrow t q: \hspace{1cm} &\frac{80}{3}g_3^4\frac{s^2+u^2}{t^2} \\
t g \rightarrow t g: \hspace{1cm} &16 g_3^4\frac{s^2+u^2}{t^2}\\
t g \rightarrow g t: \hspace{1cm} &-\frac{64}{9} g_3^4\frac{s}{u}
\end{align}

\begin{center}\uline{$SU(2)$ Gauge Boson Annihilation:} \end{center}
\vspace{-.5cm}
\begin{align}
W g\rightarrow q \bar{q}: \hspace{1cm} &24g_3^2 g_2^2\frac{u}{t} \\
W q\rightarrow q g: \hspace{1cm} &24g_3^2 g_2^2 \frac{u}{t} \\
W W\rightarrow f \bar{f}: \hspace{1cm} &\frac{9}{2}g_2^4\left(\frac{u}{t}+\frac{t}{u}\right)
\end{align}

\begin{center}\uline{$SU(2)$ Gauge Boson Scattering:} \end{center}
\vspace{-.5cm}
\begin{align}
 W f\rightarrow Wf: \hspace{1cm} &60g_2^4  \frac{s^2+u^2}{t^2} \\
 W f\rightarrow f W: \hspace{1cm} &-9 g_2^4 \frac{s}{u} .
\end{align}

It should be noted that some of the amplitudes listed above do not simplify to those found in Ref.~\cite{Moore:1995si}.  This is due to some errors in the treatment of Ref.~\cite{Moore:1995si} as was subsequently pointed out in Ref.~\cite{Arnold:2000dr}.  We have taken these into account and checked that our results for the squared matrix elements do match up to those listed in Refs.~\cite{Arnold:2002zm, Arnold:2003zc}, where applicable.

\end{document}